\documentclass[twocolumn,numberedappendix,twocolappendix,appendixfloats]{openjournal_dhayaa}

\usepackage[T1]{fontenc}
\usepackage[utf8]{inputenc}
\usepackage{comment}
\usepackage{amsmath}
\usepackage{graphicx}
\graphicspath{{./}}
\usepackage{orcidlink}
\usepackage{xcolor}
\usepackage{xspace}
\usepackage{amsmath}
\usepackage{enumitem}
\usepackage{natbib}
\setcitestyle{aysep={}}

\definecolor{xlinkcolor}{cmyk}{1,1,0,0}

\newcommand{\CHECK}[1]{\textcolor{red}{CHECK: #1}}
\newcommand{\NC}[1]{\textcolor{orange}{CITE: #1}}
\newcommand{\ca}[1]{\textcolor{purple}{CA: #1}}
\newcommand{\kb}[1]{\textcolor{blue}{KB: #1}}
\newcommand{\jgm}[1]{\textcolor{orange}{JGM: #1}}
\newcommand{\er}[1]{\textcolor{green}{ER: #1}}
\newcommand{\gm}[1]{\textcolor{red}{GM: #1}}
\newcommand{\rh}[1]{\textcolor{teal}{RH: #1}}

\newif\ifshowcomments
\showcommentstrue

\ifshowcomments
\else
  \renewcommand{\CHECK}[1]{}
  \renewcommand{\NC}[1]{}
  \renewcommand{\ca}[1]{}
  \renewcommand{\kb}[1]{}
  \renewcommand{\jgm}[1]{}
  \renewcommand{\er}[1]{}
  \renewcommand{\gm}[1]{}
  \renewcommand{\rh}[1]{}
\fi

\begin{document}
\title{Imprints of Mass Accretion History on Galaxy Cluster Morphology}

\date{July 2025}

\author{{Kabelo~Tsiane\,$^\dagger$}\orcidlink{0009-0008-6557-2065}}
\affiliation{Leinweber Institute for Theoretical Physics, University of Michigan, Ann Arbor, MI 48109, USA}
\affiliation{Department of Physics, University of Michigan, Ann Arbor, MI 48109, USA}
\author{Camille Avestruz\orcidlink{0000-0001-8868-0810}}
\affiliation{Leinweber Institute for Theoretical Physics, University of Michigan, Ann Arbor, MI 48109, USA}
\affiliation{Department of Physics, University of Michigan, Ann Arbor, MI 48109, USA}
\author{Elena Rasia\orcidlink{0000-0003-4175-002X}}
\affiliation{INAF – Osservatorio Astronomico di Trieste, Via Tiepolo 11, I-34131 Trieste, Italy}
\affiliation{Department of Physics, University of Michigan, Ann Arbor, MI 48109, USA}
\author{Roan Haggar\orcidlink{0000-0001-5490-9621}}
\affiliation{Waterloo Center for Astrophysics, University of Waterloo, Waterloo, Ontario N2L 3G1, Canada}
\affiliation{Department of Physics and Astronomy, University of Waterloo, Waterloo, Ontario N2L 3G1, Canada}
\author{Jesse B. Golden-Marx\orcidlink{0000-0002-6394-045X}}
\affiliation{School of Physics and Astronomy, University of Nottingham, Nottingham NG7 2RD, UK}
\author{Guillaume Mahler\orcidlink{0000-0003-3266-2001}}
\affiliation{STAR Institute, Quartier Agora – Allée du six Août, 19C B-4000 Liège, Belgium}
\author{Elizaveta Sazonova\orcidlink{0000-0001-6245-5121}}
\affiliation{Waterloo Center for Astrophysics, University of Waterloo, Waterloo, Ontario N2L 3G1, Canada}
\affiliation{Department of Physics and Astronomy, University of Waterloo, Waterloo, Ontario N2L 3G1, Canada}
\author{James Taylor\orcidlink{0000-0002-6639-4183}}
\affiliation{Waterloo Center for Astrophysics, University of Waterloo, Waterloo, Ontario N2L 3G1, Canada}
\affiliation{Department of Physics and Astronomy, University of Waterloo, Waterloo, Ontario N2L 3G1, Canada}
\author{Massimo Meneghetti\orcidlink{0000-0003-1225-7084}}
\affiliation{INAF – Osservatorio di Astrofisica e Scienza dello Spazio di Bologna, Via Piero Gobetti 93/3, I-40129 Bologna, Italy}

\email{$^\dagger$ kabelo@umich.edu}

\begin{abstract}
Variations in dynamical states of galaxy clusters can introduce biases and scatter in observable-mass relations.
The dynamical state of a cluster is an emergent feature of its mass accretion history (MAH), it is therefore useful to constrain the MAH of the cluster.
In this work, we characterize 305 massive clusters from The300 project by connecting features from their projected stellar distributions to their mass accretion histories (MAH).
As a baseline, we first correlate standard host dark matter halo dynamical state indicators at $z=0$ with their MAHs via the Spearman rank correlation coefficient $\rho_{\mathrm{sp}}$.
Both substructure mass fraction and center-of-mass offset measurements correlate strongly ($\rho_\mathrm{sp}\gtrsim0.5$) with the MAH measured between $0.1\lesssim z\lesssim 1$.
We repeat this exercise with various morphological measurements of projected stellar density maps, many of which exhibit moderate correlation strength with different times in the MAH.
Broadly, core morphological measurements ($r \leq 30\,\mathrm{kpc}$) correlate better with early-time MAH.
On the other hand, core-excised ($50\,\mathrm{kpc} \leq r \leq 1\,\mathrm{Mpc}$) morphological measurements correlate better with late-time MAH.
We further quantify the MAH prediction power of both traditional dynamical state indicators and morphological parameters using Multivariable Conditional Abundance Matching (MultiCAM), a framework for connecting galaxy/halo properties with their formation history.
MultiCAM employs simple rank-ordering operations, making it straightforward to translate to observed datasets.
We find reasonable ($\rho_{\mathrm{sp}} \geq 0.6$) performance for predictions of the mass fraction between $1\lesssim z\lesssim 0.1$, though with notable information loss when using projected quantities ($\rho_{\mathrm{sp}} \geq 0.4$).
In one example application of our methodology, we use the coefficients of the MultiCAM models to select subsamples of galaxy clusters that have accreted more (or less) of their $z = 0$ mass budget over a given time frame.
\end{abstract}


\maketitle

\section{Introduction}
Galaxy clusters are the largest gravitationally collapsed objects in the Universe, making them cosmological probes and astrophysical laboratories for galaxy formation.
Measurements of the composition and evolution of the abundance of galaxy clusters constrain cosmological models; clusters trace the growth of structure and energy density content of the Universe \citep{allen_cosmological_2002}.
Clusters are approximately composed of 85\% dark matter and 15\% baryonic matter in the form of the intracluster medium (ICM) and hundreds to thousands of galaxies \citep{santos_multi-wavelength_2016}.
These objects form from peaks in the matter density field in the early Universe.
As the Universe expands to suppress structure formation, gravitational collapse promotes hierarchical growth of the density peaks through accretion of matter and mergers between galaxies, galaxy groups, and galaxy clusters -- leading to the distribution of clusters we see today \citep{allen_cosmological_2002}.

These mergers uniquely define the mass accretion history (MAH) of a cluster, with the onset of each merger marking an epoch of disequilibrium leading to continuous variations in cluster dynamical state.
Clusters exhibit largely monotonic evolution of their $M_{200c}$ masses through cosmic time, either through merger events, or what is termed the "pseudo-evolution".
The pseudo-evolution describes how the diluting background density field of the Universe increases halo $M_{200c}$ masses \citep{diemer_pseudo-evolution_2013}, and therefore clusters with relatively quiescent accretion histories still exhibit smooth mass growth, and evolution of the features that describe them (e.g. halo concentration, scale radius).
As shown in \cite{wang_concentrations_2020}, the onset of a merger causes a deviation from the pseudo-evolutionary track for the halo's features, leaving an imprint that can last several dynamical times.
The dynamical time is the time taken to cross a significant fraction of an equilibrium dynamical system: $$t_d (z=0)\simeq\frac{R_{200}}{V_{\mathrm{circ}}} = \sqrt{\frac{R_{200}^3}{GM_{200}}} = \frac{1}{10H_0}\approx 1.44\,\mathrm{Gyr}$$
We associate this concept of pseudo-evolution with the process of slow accretion that maintains equilibrium; we associate merger events with fast accretion that causes the system to fall out of equilibrium.

Constraining the masses of these out-of-equilibrium systems can be challenging.
During and after a merger, the assumption of hydrostatic equilibrium breaks down which causes a bias in the hydrostatic mass estimate (see \cite{braspenning_hydrostatic_2024}).
A long-standing issue in the cluster-simulations community is the ~20\% bias in mass estimates that rely on the assumption of hydrostatic equilibrium.
Previous studies have shown weak correlations between proxies of the dynamical state of a cluster and the hydrostatic mass bias (see \cite{gianfagna_study_2022}), possibly due to the definition of the dynamical state proxy used.
Mergers have also been found to introduce bias and scatter in observable-mass scaling relations (e.g $T_X - M$, $L_X - M$) to varying degrees across different studies (\cite{planelles_galaxy_2009, krause_merger-induced_2012, chen_imprints_2019},\cite{kong_merger-induced_2026}).

Poorly constrained mass estimates propagate into cosmological parameter inference.
Cluster cosmology analyses are sensitive to $\sigma_8$, the amplitude of the matter power spectrum, and $\Omega_m$, the matter density parameter(see \cite{allen_cosmological_2002, ettori_cluster_2009, mantz_cosmological_2021, mpetha_cosmology_2025}).
For example, several studies (e.g. \cite{ade_planck_2016, salvati_mass_2019, lesci_mass_2023}) found that calibrating the mass bias on low-redshift (high-redshift) datasets leads to smaller (larger) bias that in turn lowers (raises) the value of $\sigma_8$ -- however the effect on $\Omega_m$ is consistently less pronounced.
In addition to the mass, intrinsic cluster properties such as triaxiality, morphology and thermodynamic profiles are affected by mergers.
Numerous studies (\cite{gallo_characterising_2024, saxena_chex-mate_2025, srinivasan_brightest_2026}) have explored how these properties (as well as observed orientation) affect survey completeness.
Thus, constraining the dynamical state is important for characterizing individual clusters as well as populations of clusters for use in cosmological studies.
The effect of dynamical state on cosmology is not uniform -- it should be emphasized that different measures of dynamical state lead to differing subsample selections (see \cite{cui_three_2018, astudillo_effect_2025-1}).

At the astrophysical level, understanding the dynamical state of clusters presents several exciting avenues of study.
Dynamically disturbed systems are unique laboratories for high-energy astrophysical phenomena such as ICM shocks, turbulent gas flows and magnetic field interactions.
Improved quantification of dynamical state indicators can help distinguish between variations in dynamical state and potential signatures of non-standard dark matter interactions; both affect the phase space distribution of the cluster dark matter halo, ICM, and stellar content.
Previous studies have found that the BCG and ICM components are affected by the merger history of the cluster (\cite{donahue_morphologies_2016, nurgaliev_testing_2017, contreras-santos_three_2022}).
The brightest cluster member is usually found near the center of the potential: the stellar content of accreted systems is stripped from the incoming halo due to dynamical friction, and it slowly descends to the bottom of the potential well where it may grow the BCG.
The diffuse gas component traces the underlying cluster potential (\cite{lau_shapes_2011, zemp_determining_2011}), therefore for a cluster experiencing a merger we may see the ICM deviate from a smooth, continuous, spherical distribution (\cite{chen_imprints_2019}) that reflects a self-gravitating system in hydrostatic equilibrium.
With a better understanding of the cluster dynamical state, we can thus probe the assembly histories of these components in greater detail.

A plethora of proxies for the cluster dynamical state have been proposed to quantify how close a galaxy cluster is to dynamical equilibrium.
They can broadly be split into two groups: theoretical and observable parameters.
Most theoretical parameters of the dynamical state are measured directly from the full 3-dimensional phase space information in simulations, and are typically not directly observable.
These include: the virial ratio, center of mass offset, substructure mass fraction (\cite{neto_statistics_2007, cui_dynamical_2017, valles-perez_choice_2023}); sparsity, velocity dispersion deviation and stellar mass gap (\cite{kim_new_2024}); spin parameter (\cite{bullock_universal_2001, valles-perez_eventful_2025}), connectivity, and accretion rate (\cite{haggar_reconsidering_2024}); the splashback radius, and the halo concentration (\cite{more_splashback_2015, diemer_universal_2015, mendoza_multicam_2023})
Other parameters leverage the MAH directly such as the time when the halo reaches half its present-day mass (often called the 'formation time'), and the mass accreted since $z=1$ or over a dynamical time.
These simulation-based metrics serve as a baseline to compare observational metrics where perfect phase space information is absent.

For observed galaxy clusters, numerous empirical metrics have been developed as proxies for the dynamical state characterization.
In the X-ray there is asymmetry and surface brightness fluctuations (\cite{zhang_locuss_2010}); centroid shift, power ratio, and concentration index (\cite{jeltema_cluster_2008, yuan_dynamical_2020}).
Each of these have found varying success across the literature in parameterizing the dynamical state.
A study by \cite{rasia_x-ray_2013} found that the third-order power ratio, the asymmetry parameter, and the surface brightness concentration best separate clusters into 'relaxed' and 'disturbed' samples when compared to 60 visually classified Chandra-like images.
A separate study by \cite{mantz_cosmology_2015} looked at 361 galaxy clusters identified by Chandra and ROSAT, classifying them into 'relaxed' and 'unrelaxed' based on X-ray morphologies -- symmetry, peakiness and alignment.
These X-ray indicators have been combined to select particularly relaxed subsamples (\cite{allen_cosmological_2002, ettori_cluster_2009, mantz_cosmological_2021}), and to account for merger driven biases in cluster mass estimation ( \cite{rasia_systematics_2006, mahdavi_evidence_2008, becker_accuracy_2011, rasia_lensing_2012, nelson_evolution_2012, battaglia_cluster_2013, nelson_weighing_2014}).
Unfortunately, sample sizes in these studies are relatively small because obtaining enough X-ray pointings with sufficient depth to make these measurements reliably remains a challenge.

Optical measures of dynamical state present a separate, more accessible avenue for characterizing clusters.
For example, the fraction of the cluster's light contained in the intracluster light (ICL).
A study conducted by \cite{golden-marx_hierarchical_2025} demonstrated that larger ICL fractions are related to relaxed systems.
Other optical measures are early-type galaxy fraction, and the difference in magnitude between the brightest and fourth brightest cluster member \cite{zarattini_fossil_2015, contini_origin_2021, brough_preparing_2024}.
There has been consistency across studies in applying these optical measures to simulations, however, there are contradictions when applied to observations.
For example, \cite{dupke_independent_2022, jimenez-teja_evidence_2024} find that a larger ICL fraction corresponds to a dynamically disturbed system, whereas \cite{montes_intracluster_2018, ragusa_does_2023} find that a larger ICL fraction corresponds to a dynamically relaxed system.
This discrepancy may exist due to how different author's identify the ICL from different datasets, but more work must be done to assess this (see \cite{brough_preparing_2024}).

As has been highlighted, characterizations of the cluster dynamical state varies across the literature.
This characterization often relies on cuts on single or combined parameters (e.g. \cite{sanders_srgerosita_2025}) or non-linear machine learning approaches (e.g. \cite{haggar_reconsidering_2024}, \cite{soltis_multiwavelength_2025}).
Note, one particular challenge in machine learning approaches lies in domain adaptation from simulations to observations (see \cite{ntampaka_importance_2025}).
Nevertheless, across the literature it is clear that the problem domain is not a simple dichotomy of having 'relaxed' or 'disturbed' clusters.
The dynamical state is more so a continuum of states, and whatever cutoffs are applied to single/multi-parameter metics is highly dependent on the end goal of the application e.g. sample selection, maximizing hydrostatic mass bias etc.
Furthermore, different metrics of dynamical classification may be sensitive to different parts of a galaxy cluster's mass accretion history.
Therefore, a cluster may be defined as relaxed with respect to a particular set of metrics, but disturbed with respect to another (see \cite{srinivasan_brightest_2026} for recent study on different indicators of assembly bias inferring different assembly histories for the cluster sample considered).
There is no consensus on how to quantify the dynamical state of individual clusters nor how we might account for variations in the dynamical state across a sample of clusters.

\begin{figure*}[tbp]
    \centering
    \includegraphics[width=0.9\textwidth]{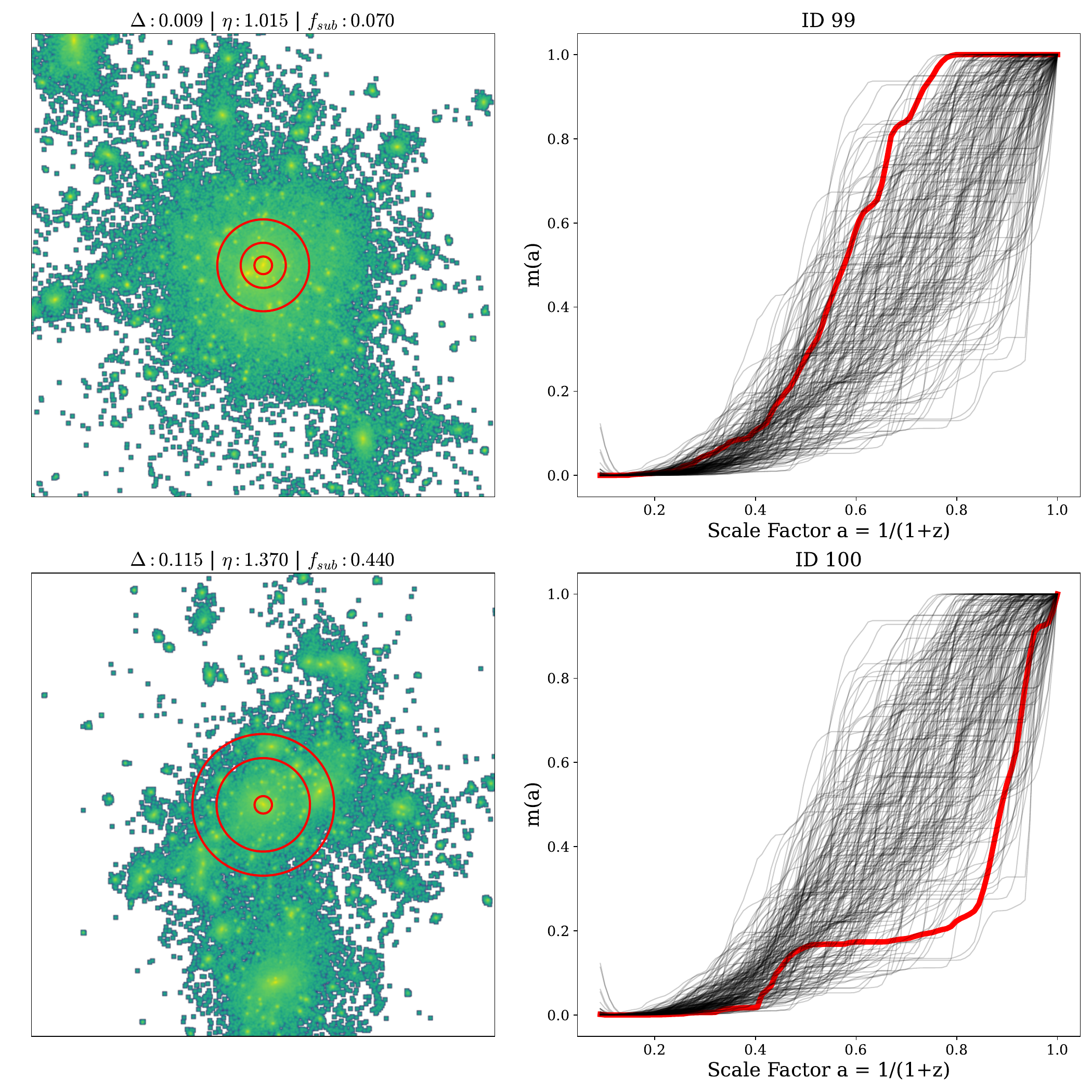}
    \caption{\textit{Example galaxy clusters from \texttt{The300} and their mass accretion histories.} Stellar density maps at $a = 0.94$ of the region surrounding an example early forming galaxy cluster (top left) and one late forming galaxy cluster (bottom left), with $a = 0.94$ dynamical state classification defined by: center of mass offset ($\Delta$), virial ratio ($\eta$), and substructure mass fraction ($f_\mathrm{sub}$) (Sec.~\ref{sec:methods:dsparams}).
    Overplotted are $r_{20}$, $r_{50}$, and $r_{80}$ (Sec.~\ref{sec:methods:statmorph}) as measured in an aperture of $50\,\mathrm{kpc} \leq r \leq 1\,\mathrm{Mpc}$.
    The panels on the right show our sample's mass accretion histories, defined as $m(a)=M_\mathrm{peak}(a)/M_\mathrm{peak}(a=1)$. We highlight the specific example cluster in the left-hand side in red.}
    \label{fig:maps}
\end{figure*}

In this work, we quantify the connection between morphological features in projected stellar density maps of simulated galaxy clusters with their host halo mass accretion histories.
We employ a galaxy cluster-halo connection approach, \texttt{MultiCAM}, that is based on rank ordering principles of abundance matching -- making the approach potentially transferable to observations.
In Sec.~\ref{sec:methods}, we describe the simulated dataset, the choices of theoretical parameters as our baseline for comparison, the morphological parameters that we measure, how we parameterize MAH, and explain the \texttt{MultiCAM} framework.
We present our results in Sec. \ref{sec:Results} where we demonstrate the imprints of mass accretion history on the parameters, quantifying their connections with predictive models of the mass accretion history.
We further provide an example application of these results in sample selection by ranked mass accretion history tracers.
We conclude in Sec.~\ref{sec:conlusion}.

\section{Methodology}\label{sec:methods}

To characterize the dynamical state of a galaxy cluster we measure morphological features using stellar density maps produced from simulated clusters in The Three Hundred dataset \citep{cui_three_2018}.
Morphological parameters are calculated using the Python library \texttt{statmorph} \citep{rodriguez-gomez_optical_2019}. We connect the mass accretion histories of our cluster sample sample with their morphological parameters using \texttt{MultiCAM}, a modeling approach for galaxy-halo connection presented in \citet{mendoza_multicam_2023} that extends the abundance matching and conditional abundance matching approaches to multivariate predictions.

\subsection{The Three Hundred Dataset}\label{sec:methods:sims}

The Three Hundred project \citep{cui_three_2018} is a collection of 324 galaxy clusters.
The dataset is created from the dark-matter-only MultiDark Planck 2 (MDPL2) \citep{klypin_multidark_2016} simulations, whose cosmology comes from the Planck mission \citep{ade_planck_2016}: $\Omega_M = 0.307,\sigma_8 = 0.823, H_0=67.8\,\mathrm{kms^{-1}Mpc^{-1}}$.
MDPL2 is a periodic cube of comoving side length $1\,\mathrm{h^{-1}Gpc}$, containing $3840^3$ DM particles of mass $1.5\times10^9\,\mathrm{h^{-1}M_{\odot}}$ each.

The 324 most massive spherical regions, with radius $r = 15\,\mathrm{h^{-1}Mpc}$, are identified in MDPL2 using the \texttt{ROCKSTAR} halo finder \citep{behroozi_rockstar_2012} at $z = 0$.
The halo virial radius is set at $\sim R_{98c}$, where the average halo density is 98 times the critical density of the Universe at $z=0$.
Note that the reference density used enforces the pseudo-evolution of halo masses as part of the MAH.
The corresponding halo virial masses are $M_{\text{vir}} \gtrapprox 8 \times 10^{14} h^{-1} M_{\odot}$.
The size of the re-simulated region is much larger than the cluster's virial radius, therefore each region contains filamentary structures and galaxy groups.
Hydrodynamical re-simulations use the Smooth-Particle-Hydrodynamics code \texttt{GADGET-X}.
The dataset includes 128 fixed snapshots, ranging from $z = 17$ to $z = 0$.
The code implements various sub-grid physical models for baryonic feedback such as AGN feedback, stellar feedback, gas mass-loss etc. (see \cite{cui_three_2018}) and references therein for a full discussion of the implemented astrophysics).
Following re-simulation the final cluster masses are $M_{200c} > 6.42 \times 10^{14} h^{-1}\text{M}_{\odot}$ at $z = 0$, corresponding to the most massive observed clusters in our Universe.

Halos in the re-simulated regions are identified using the \texttt{Amiga Halo Finder} (AHF) code \citep{knollmann_ahf_2009}, which includes the stellar and gas components.
Halo properties are calculated within a spherical region of radius $R_{200c}$.
Subhalos are then identified as those halos found within $R_{200c}$.
Subhalo properties (e.g. subhalo mass) are calculated using a truncated radius $R_t < R_{200c}$ that is marked by a characteristic upturn in the density profile, where $R_{200c}$ would be their own actual radius if found in isolation.

\subsubsection{Stellar Density Maps}
We use stellar density maps for 305  galaxy clusters to extract morphological features to connect to MAH.
For each cluster we consider 3 projections along 3 orthogonal lines of sight.
The stellar density maps are generated at snapshot 125 ($a=0.94$ or $z=0.06$) with the code described in \cite{meneghetti_introduction_2021}, which has been used both to reproduce stellar density maps and mock weak lensing catalogs \citep{meneghetti_excess_2020, meneghetti_probability_2022, meneghetti_persistent_2023}.
The maps have a side length of 5 Mpc with 2048 pixels and consider an integration along the line of sight of 10 Mpc.

Fig.~\ref{fig:maps} shows two examples of an early forming and late forming system.
The left panels show the stellar density maps of these systems at $a=0.94$ ($z=0.06$).
Here we loosely define early forming as having accumulated half of its present-day mass before $a<0.5$ ($z>1$), and late forming as reaching half of its present-day mass after $a > 0.7$ ($z<0.43$).
The thin black lines in the right panels show the MAH of all systems, with the MAH of the specific examples highlighted in the thick red line.
We describe our mass accretion history parametrization in further detail in Sec.~\ref{sec:methods:mah}.
From the example maps, we see that the early forming cluster is the more spherically symmetric one, as expected for a relatively virialized system.
We annotate each example with their dynamical state parameters, which we describe in the next subsection.

\subsection{Dynamical State Parameters}\label{sec:methods:dsparams}
As a benchmark for comparison, we adopt the following 3D dynamical state parameterizations of a dark matter halo as prescribed by \cite{neto_statistics_2007} and \cite{cui_dynamical_2017}:
\begin{itemize}
    \item Substructure mass fraction $f_m$: the mass fraction in resolved subhalos $f_m = \sum M_{\text{sub}} / M_{200c}$, where $M_{\text{sub}}$ is the mass of each subhalo.
    \item Center of mass offset $\Delta$: the offset between the center of mass of the cluster and the density peak of the halo $\Delta = |R_{\mathrm{cm}} - R_\mathrm{c}| / R_{200c}$
    \item Virial ratio $\eta$: the ratio of kinetic energy, $T$, to potential energy, $W$, of the cluster, modified by the energy from surface pressure $E_s$ $\eta = (2T-E_s) /|W|$
\end{itemize}
Larger values of $f_m$ suggest that there are more recently accreted subhalos that have not been in the host halo long enough to be stripped and/or accreted onto the massive central, and is therefore a more dynamically disturbed halo.
Similarly for $\Delta$, a larger offset indicates a more dynamically disturbed halo; mergers are anisotropic events that shift the center of mass towards the accreting body whilst not shifting the density peak.
Finally, $\eta$ measures the departure of the system from virial equilibrium, where a larger value indicates a more disturbed halo.
The extra term $E_s$ takes into account that dark matter halos are not isolated systems as they exist in cosmic environments including filaments, voids, neighboring clusters, etc. (see \cite{davis_virialization_2011} for a full treatment of the virial theorem on dark matter halos).
All parameters are calculated at $R_{200c}$ and $R_{500c}$.
The main results in Sec.~\ref{sec:Results} are for dynamical state measurements made within $R_{200c}$; for the same results at $R_{500c}$ we observe broadly the same trends, albeit weaker correlations when using the smaller aperture (see Fig.~\ref{fig:mahdscorr500c}).
These results suggests that the MAH signal is best constrained using the outskirts of the cluster.

These three parameters have been used in the literature to classify the dynamical state of simulated dark matter host halos, and have been applied to \texttt{The300} simulated galaxy cluster sample used in this work \citet{cui_dynamical_2017}.
Note, it is difficult to estimate such definitions of dynamical state parameters from projected luminous components.
We therefore use these dynamical state measurements as a benchmark for connections to the mass accretion history that we compare to projected measurements of the stellar component of galaxy clusters using morphological measurements with \texttt{statmorph} described in the next section.

\subsection{Observables/Morphological Measurements}\label{sec:methods:statmorph}

The relevant observable signatures are the 2D morphologies measured from the cluster's stellar density maps, and the magnitude gap, $m_{14}$.
$m_{14}$ is the apparent magnitude difference between the brightest cluster galaxy (BCG, measured within 30 kpc) and fourth brightest cluster member within $0.5R_{200c}$ \citep{dariush_mass_2010}.
This choice for the outer radius for the BCG is largely context dependent.
It was found by \cite{zhang_dark_2024} that the luminosity of the total stellar component measured at $r<30\,\mathrm{kpc}$ is dominated by the diffuse stellar component (CG + ICL).
For both total and diffuse components the evolution of luminosity with richness, a proxy for total cluster mass, was found to be more pronounced in apertures $r>50\,\mathrm{kpc}$ compared to the $r<30\,\mathrm{kpc}$ aperture.
The ICL is the only part of the diffuse component in this $r>50\,\mathrm{kpc}$ region, therefore these two observations justify identifying the the central galaxy with outer radius $r<30\,\mathrm{kpc}$.

We use the python library \texttt{statmorph} \citep{rodriguez-gomez_optical_2019}, for calculating non-parametric morphological diagnostics (e.g., Gini-M$_{20}$ \citep{lotz_new_2004} and concentration-asymmetry–smoothness statistics \citep{conselice_relationship_2003}), as well as fitting 2D Sérsic profiles.
\texttt{statmorph} was originally intended for use with galaxy images, but its flexibility allows us to apply it to stellar density maps of galaxy clusters, especially for the features we care about in this study -- half-light radius, concentration, asymmetry and Sérsic amplitude.
There is one key difference in our \texttt{statmorph} pipeline compared to a normal pipeline that deals with the segmentation map.
The segmentation map labels pixels belonging to different sources, usually detected using specialized tools.
The present study uses images of each cluster directly, with no foreground/background structures to contend with, nor any PSF applied.
Therefore the segmentation is map is made by simply creating a circular/annular mask centered on the image center. We label everything inside the mask as one source, and everything outside as the background which has value 0.
Given we are working with noiseless simulation data, having a background of 0 is appropriate.
This choice of segmentation map affects how total flux is calculated.
\texttt{statmorph} defines the total flux of a source as the total flux within $1.5r_{\text{petro}}$, the Pertrosian radius \citep{petrosian_surface_1976}.
$r_\text{petro}$, which is the radius where the mean surface brightness is equal to 20\% of the surface brightness at that radius.
Our choice of segmentation map makes $r_\text{petro}$ ill-defined, and so the actual aperture that is being used to calculate total flux will differ slightly.

Morphological features are measured in an annular aperture with $50 \mathrm{kpc} \leq r \leq 1 \mathrm{Mpc}$, and a circular aperture $r \leq 30 \mathrm{kpc}$.
These apertures were chosen because they gave the strongest correlations (see Appendix \ref{sec:app:corevsannuluscorrelation} for more details).

Half-light radius is the radius of a circular aperture at which half the total light from the source is enclosed.
Concentration is defined as:
\begin{equation}\label{eq:3}
    C = 5\text{log}_{10}\left(\frac{r_{80}}{r_{20}}\right),
\end{equation}
where $r_{20}$ and $r_{80}$ are the radii of circular apertures containing 20\% and 80\% of the source light.
The asymmetry index (A) is calculated by subtracting the image rotated by 180º from the original image:
\begin{equation}
    A = \frac{\sum_{i, j}|I_{ij} - I_{ij}^{180}|}{\sum_{i, j}|I_{ij}|},
\end{equation}
where $I_{ij}$ and $I_{ij}^{180}$ are the pixel values of the original and rotated image respectively.
For all of these quantities, the center of the aperture/image is always determined by the point that minimizes $A$.
Finally, the Sérsic amplitude is simply the intensity at $r=0$ of the Sérsic profile:
\begin{equation}
    I(r) = I_0\text{exp}\left(-b_n\left[\frac{r}{r_{50}}\right]^{1/n} - 1)\right)
\end{equation}
where $r$ is the radius, $r_{50}$ is the half-light radius, $n$ is the Sérsic index and $b_n$ is a constant solved for numerically.
The Sérsic profile poorly models the clumpier flux distribution of a galaxy cluster (as opposed to the smoother stellar component of a galaxy), and indeed our choice to include the Sérsic amplitude is motivated \textit{a posteriori} from results which we discuss in Section \ref{sec:Results}
We perform these measurements using circular and annular apertures at varying radii in order to probe the different regions of the cluster that hold information on different eras of the mass accretion history.

We also include information on the magnitude gap $m_{14}$, which we take as the mass ratio between the most massive, and fourth most massive system within $0.5R_{200c}$.
Historically, this parameter has been used to trace the BCG's growth \citep{golden-marx_impact_2018, golden-marx_hierarchical_2025} based on the hierarchical growth model given by the $\Lambda\text{CDM}$ model, and supported by observations, simulations and semi-analytic models.
At high redshift ($z>2$) a dense core ($r\lessapprox10$kpc) containing $\approx 25\%$ of the cluster's stellar mass forms.
This is the seed for the BCG, which is measured within a 30 kpc outer radius.
N-body simulations show that the BCG growth is positively correlated with the cluster's growth \citep{solanes_forming_2016}, while other cluster member's remain stagnant.
Not much growth is seen in the other cluster members because the accretion of smaller, less-luminous progenitor galaxies in the outer envelope does little to affect the apparent magnitude of the 4th brightest cluster member within $0.5R_{200c}$.
However, over time the in-fall of these new systems into the center of the cluster leads to BCG growth.
Thus, for systems that have experienced more merger activity in the past, one expects a larger measured m14.
In other words, for a dynamically relaxed system (merger activity is not recent) we expect a larger m14.

\subsection{Mass Accretion History}\label{sec:methods:mah}

The mass accretion history (MAH) of a dark matter halo often refers to how the mainline progenitor of a host dark matter halo accumulates mass over time.
Both smooth accretion of matter and the merger of smaller dark matter halos contribute to the MAH.
In our case, we trace the MAH of the mainline progenitors at $z=0$ to connect these with indicators of present-day dynamical state.
We expect clusters that have recently (within several dynamical times $t_d \approx 1.44\,\mathrm{Gyr}$ experienced or are experiencing a major merger to be dynamically disturbed.
By utilizing the entire MAH, we aim to encode a more nuanced description of the present-day dynamical state.

The MAH has several parameterizations across studies; for this analysis we focus on one, with results for two more presented in Sec. \ref{sec:app:mah}.
First we calculate the normalized peak mass as a function of scale factor,
\begin{equation}\label{eq:norm}
    m(a) = \frac{M_{peak}(a)}{M_{peak}(a=1)}
\end{equation}
where
\begin{equation}\label{eq:mpeak}
    M_{\text{peak}}(a) = \max_{0 \leq a' \leq a} \left[ M_{\text{vir}}(a') \right]
\end{equation}

The normalization here enforces monotonicity of the function, which is a valid treatment for these massive systems.
The difference between $M_{\text{peak}}(a)$ and $M_{\text{vir}}(a)$ is small for the main progenitor halos in The300, thus their mass growth is typically monotonically increasing.
We note that this approximation would fail for subhalos (and their associated stellar and gas populations) due to a more significant impact from baryonic feedback and tidal stripping \citep[e.g.][]{wechsler_connection_2018}.
Many systems have missing values at early-to-intermediary snapshots.
To account for this we use a piecewise cubic Hermite interpolating polynomial to calculate $m(a)$ for each system, mapped onto a common set of scale factors while preserving the monotonicity and shape of the growth curve.

We then connect the MAH to both the dynamical state (DS) measurements described in Sec.~\ref{sec:methods:dsparams} and the \texttt{statmorph} measurements described in Sec.~\ref{sec:methods:statmorph} using the galaxy-halo connection model, \texttt{MultiCAM}, described in the next subsection.

\subsection{MultiCAM}\label{sec:multicam}

MultiCAM \citep{mendoza_multicam_2023} is a Multi-variable Conditional Abundance Matching framework based on the same principles as traditional abundance matching \cite{kravtsov_tumultuous_2004} and conditional abundance matching.
Typically, abundance matching is used in galaxy-halo connection modeling; but here we are leveraging the tool to quantify how morphologies and dynamical states are connected with MAH.
Historically, galaxy/halo properties are connected to MAH using a single-parameter summary of the MAH (in this study we use $m(a)$) or the value returned by a single-parameter fit.
This leads to a one-to-one parameter correlation analysis -- abundance matching.
The abundance matching algorithm is built off of one simple assumption -- perfect correlation between the two parameters in the model.
The algorithm evaluates the $N_\star^{-1}(N_{\text{dm}}(M_{\text{vir}}))$, where $N_\star$ is the observed cumulative stellar mass function, and $N_{\text{dm}}$ is the theoretical cumulative mass function.
Despite its simplicity, the method has found success in galaxy-halo connection modeling by assigning stellar masses/luminosities to simulated halos.

A hierarchical extension to abundance matching, Conditional Abundance Matching (CAM) \citep{hearin_dark_2014}, is primarily used to generate empirical models from observable properties.
It optimally implements the assumption that a given target, $Y$, is monotonically determined by a given feature, $X$.
In this work the targets are the normalized peak mass, $Y_m$, the features are the morphologies, $X_{\mathrm{SM}}$ and the dynamical state parameters, $X_{\mathrm{DS}}$.
CAM assigns the normalized peak mass by evaluating $Y_m = F_m^{-1}(F_{\text{DS}}(X_{\text{DS}})|a)$, where $F_{\text{DS}}$ is the dynamical state CDF (or the CDF of morphology), and $F_{\text{m}}$ is the CDF for the targeted normalized peak mass at a fixed scale factor, $a$.
In other words: given a measured dynamical state (morphology) at $a = 1$, CAM assigns the normalized peak mass at some scale factor, $m(a)$.

MultiCAM is a novel algorithm that extends the CAM framework to predict multiple targets from multiple features.
The algorithm works as follows:
\begin{enumerate}
    \item Marginally transform each feature and target from their empirical distribution to a Gaussian distribution using an inverse transform method (e.g. \cite{devroye_sample-based_1986} $\boldsymbol{X}, \boldsymbol{Y}\rightarrow \boldsymbol{\tilde{X}}, \boldsymbol{\tilde{Y}}$. The transformations will necessarily be different for each feature and target.
    \item Train a linear regression model for prediction in this Gaussianized space for the marginally transformed $\boldsymbol{\tilde{X}}_\mathrm{train}$ and $\boldsymbol{\tilde{Y}}_\mathrm{train}$.
    \item Use the marginally transformed test set features, $\boldsymbol{\tilde{X}}_\mathrm{test}$ to predict targets, $\boldsymbol{\tilde{Y}}_\mathrm{pred}$.
    \item Apply another quantile transformation to make the predicted distributions marginally Gaussian as well, such that it matches the $\boldsymbol{\tilde{Y}}_\mathrm{train}$ distribution.
    This is the abundance matching step between predictions and transformed targets.
    \item Perform the inverse of the quantile transformation used to map $\boldsymbol{Y}_{\mathrm{train}} \rightarrow \boldsymbol{\tilde{Y}}_\mathrm{train}$ to produce the final MultiCAM predictions: $\boldsymbol{\tilde{Y}}_\mathrm{pred} \rightarrow \boldsymbol{Y}_{\mathrm{pred}}$
\end{enumerate}

In this work we use different sets of the dynamical state proxies (features) to predict the full MAH (targets) from $a \in (0.05, 1]$ ($z \in (19, 0]$).

\section{Results}\label{sec:Results}
The mass accretion history of a galaxy cluster leaves imprints on indicators of its dynamical state, including its morphological features.
In the following subsections, we connect the mass accretion histories of the main progenitor to both theoretical dynamical state (DS) parameters, and morphological \texttt{statmorph} (SM) parameters on the projected density maps.
We first provide an initial quantification of the strength of the imprints through the Spearman rank correlation coefficient ($\rho_\mathrm{sp}$) between the mass accretion histories and the evolution of the DS parameters.
$\rho_\mathrm{sp}$ indicates which parameters have more of a monotonic relationship with the mass fraction, $m(a)=M(a)/M(a=1)$, at any given epoch (Sec.~\ref{sec:results:smpredmah}\footnote{We have also quantified these with respect to $a(m)$, the epoch at which the main progenitors reach different mass fractions see Sec. \ref{sec:app:mah}}).
$\rho_\mathrm{sp} = 1$ means the variables have an identical rank, and $\rho_\mathrm{sp}=-1$ means the variables have fully opposed ranks.
We utilize the \texttt{MultiCAM} framework to assess the predictive power of the DS and SM parameters measured at the present day, $a=1$ ($z=0$), on the overall mass accretion history in Sec.~\ref{sec:results:dspredmah} and Sec.~\ref{sec:results:smpredmah}.
Finally, as an example MultiCAM application we utilize the coefficients learned by the underlying linear regression model to illustrate the potential use of our model for sample selection by fast or slow accreting clusters defined at different epochs (Sec~\ref{sec:results:coeffselectsample}).

\subsection{Imprints of MAH on dynamical state parameters over cosmic time}\label{sec:results:mahdscorr}

\begin{figure*}[tbp]
    \includegraphics[width=0.9\textwidth]{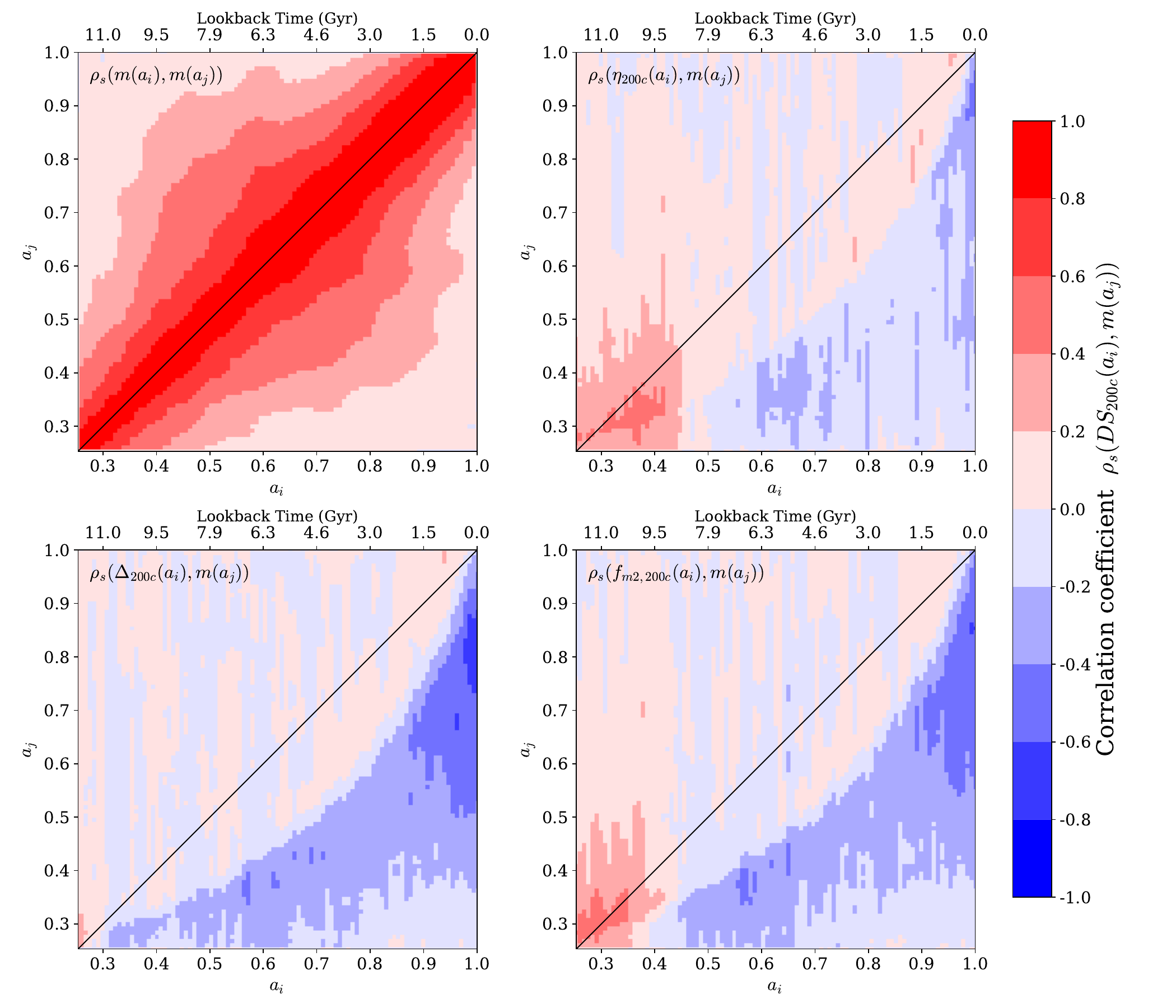}
    \caption{\textit{Correlation between the mass accretion history, $m(a_j)$, and the evolution of traditional dynamical state parameters measured at $R200c$, $DS_{200c}(a_i)$.}
    The colors in each pixel represent the Spearman correlation between $DS_{200c}(a_i)$ and $m(a_j)$, each at times $a_i$ and $a_j$ respectively, with the exception of the top left panel showing the Spearman correlation between mass accreted at different times, $m(a_i)$ and $m(a_j)$.
    The diagonal, over-plotted in black, corresponds to measurements taken at the same time, $a_i = a_j = a$.
    (a) Mass accretion at any epoch is tightly correlated with past and future growth ($|\rho_s|\gtrsim0.8$).
    (b) $\eta_{200c}$ measured at $a_i \sim 1$ has moderate-to-weak correlation strength with $m(0.85\lesssim a_j \lesssim 0.95)$
    (c) $\Delta_{200c}$ measurements from $0.9\lesssim a_i \lesssim 1$ trace mass accreted between $0.55\lesssim a_j\lesssim0.95$ with correlation strengths more than $|\rho_s|\gtrsim0.4$, with a brief peak where $|\rho_s|\gtrsim0.6$.
    (d) $f_{m,200c}$ measurements from $0.9\lesssim a_i \lesssim 1$ trace mass accretion over a similar time frame with similar correlation strength.
}\label{fig:mahdscorr}
\end{figure*}

The imprint of an accretion event cannot exist for the entire lifetime of a galaxy cluster, nor is it necessarily immediately detectable.
In Fig.~\ref{fig:mahdscorr} we illustrate the characteristic time lag of the signal from a major accretion event in terms of which epoch of the MAH it traces (see Fig~\ref{fig:mahdscorr500c} for analogous result for parameters measured within $R_{500c}$).
These heatmaps illustrate the strength of the Spearman rank correlation coefficient between the MAH defined in Sec.~\ref{sec:methods:mah}, and the time-evolution of each dynamical state (DS) parameter, described in Sec.~\ref{sec:methods:dsparams}.
For a given measurement of DS at some $a_i$ bin, we interpret the time lag as the position below the black diagonal (where $a_i = a_j$) with the largest $|\rho_{\mathrm{sp}}|$.

The top left panel (a) shows the internal correlation strength of the mass accretion histories, similar to Fig.~3 from \citet{mendoza_multicam_2023}, which measured the same internal correlation for lower mass halos ($M_{vir}\approx10^{12}h^{-1}M_\odot$.
Note that there are very strong correlations of $\rho_s(m(a_i),m(a_j))\gtrsim0.8$ for $a_i\approx a_j \pm 0.1$, indicating that the mass accretion at any given point in time is tightly correlated with mass accretion occurring before and after within a dynamical time.
These accretion events should therefore leave their imprints on these parameters for several dynamical times, particularly since there would be a delay between accretion or merger events and different signatures in the dark matter halo (e.g. \cite{wang_concentrations_2020} and \cite{wang_merger_2025}).
Note that the correlation in the latest times of scale factor is somewhat artificial because all clusters have $m(a = 1)\equiv 1$ by definition, while most systems likely have $m(a)=1$ for some preceding epochs when they first reach their peak mass.

All DS parameters are calculated at $R_{200c}$ in Fig.~\ref{fig:mahdscorr}.
In Fig.~\ref{fig:mahdscorr500c}, we show the analogous figure for DS measurements at $R500c$, where we observe broadly the same trends, albeit with much shorter peak correlations.
Normalized peak mass at $a_j$ is correlated with the following parameters measured at $a_i$: virial ratio $\eta_{200c}$ in the top right panel (b), center of mass offset, $\delta_{200c}$ in the bottom left panel (c), substructure fraction, $f_{m,200c}$ in the bottom right panel (d).
The solid black diagonal line indicates where $a_i = a_j$.
The delayed anti-correlation in each panel indicates that galaxy clusters with the lowest normalized mass (i.e. the relatively "late-forming" objects) will have larger DS parameter values later on, consistent with the physical interpretations of all these parameters in virialized systems.

To talk through a specific example, the bottom right panel illustrates that for the substructure mass fraction measured today, $f_{m, 200c}(a_i = 1)$, there is a strong anti-correlation ($0.6 \leq |\rho_{\mathrm{sp}}| \leq 0.8$), with normalized peak mass measured between $0.65 \lesssim a_j \lesssim 0.95$ ($0.54 \gtrsim z \gtrsim 0.05 $).
In other words, the information of the mass fraction of clusters measured as early as $a \approx 0.65$ ($z\approx 0.54$) leaves its imprint on the substructure mass fraction calculated today.
The center of mass offset measured today ($\Delta_{200c}(a_i =1)$) shows a systematic correlation over a similar time period, albeit weaker ($0.4 \leq |\rho_{\mathrm{sp}}| \leq 0.6$).

Finally, the virial ratio $\eta_{200c}(a_i=1)$ shows an even weaker correlation with mass accretion history.
The trend is still systematic with a weak-to-moderate anti-correlation strength ($0.4 \leq |\rho_{\mathrm{sp}}| \leq 0.6$) with $m(a_j\approx 0.9$).
These results indicate that mass accretion leaves the strongest imprint on $f_{m, 200c}$ , followed by $\Delta_{200c}$  and finally $\eta_{200c}$ measured at $a_i=1$ ($z=0$).
The cause of the blue arc in each panel is unclear, we discuss it briefly in Sec.~\ref{sec:conlusion}.

At earlier times ($a\sim 0.35$) we find that $\eta_{200c}$ and $f_{m, 200c}$ are both moderately, positively correlated with the peak mass.
This suggests that the most rapid growth of these young systems is governed by many minor mergers with small progenitors in a denser cosmic environment, leading to larger DS parameter values measured at approximately the same time.  In other words, the galaxy clusters whose main progenitor line had larger mass fractions at $0.3\lesssim a_j\lesssim 0.4$, i.e. relatively early forming objects, also have virial ratio and substructure mass fraction values indicating their more dynamically unrelaxed state at the same time.  Interestingly, the center of mass offset continues to exhibit a time delayed response even for offsets measured at $a_i\approx 0.3$.
More investigation is needed to explore these early time trends in closer detail.  We focus our subsequent results and discussions on imprints measurable at $z=0$.

\subsection{Predictive power of dynamical state parameters on mass accretion history}\label{sec:results:dspredmah}
The previous section illustrated the imprints of MAH on the evolution of dynamical state parameters.
Now, we test the prediction power of only the present-day ($z=0$) dynamical state parameters for predictions of the entire MAH.
This is simply creating a higher resolution illustration of the correlations in the final $a_i$ bin from Fig. \ref{fig:mahdscorr}.
As described in section \ref{sec:methods:mah}, the 3D dynamical state parameters in our analysis are quantities only directly measurable in simulations.
Motivated by \cite{golden-marx_hierarchical_2025}, \cite{dariush_mass_2010}, \cite{hearin_mind_2013}, we also include the 3-dimensional magnitude gap parameter, $m_{14,3D}$.
The 3-dimensional magnitude gap parameter is the mass ratio between the fourth most massive and the most massive cluster member galaxy identified within $0.5R_{200c}$.

\begin{figure*}[ht]
    \centering
    \includegraphics[width=\textwidth]{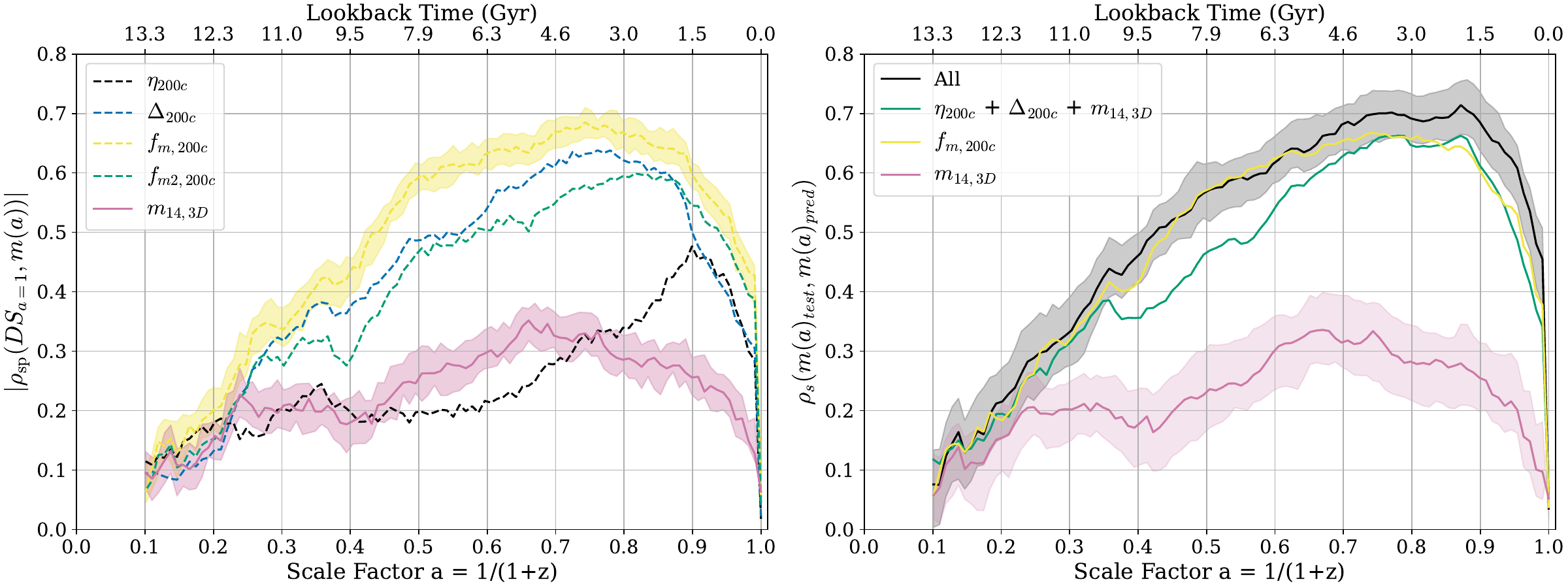}
    \caption{\textit{Left: Correlation between mass accretion history and traditional dynamical state parameters}. Solid lines show the Spearman correlation between the mass fraction at a given scale factor, $M(a)/M(a=1) = m(a)$, and various dynamical state parameters from the dark matter particles measured at $a=1$ in addition to the magnitude gap $m_{14}$ measured in 3-dimensions. Dashed lines show parameters that negatively correlate with $m(a)$. Shaded regions show example 25th-75th percentile widths of a bootstrap.  \textit{Right: Relative prediction power of traditional dynamical state parameters with \texttt{MultiCAM}} Solid lines show the Spearman correlation between the the true mass fraction at a given scale factor and mass fraction predicted using different subsets of dynamical state parameter measurements as features for \texttt{MultiCAM}. Shaded regions show example 25th-75th percentile widths. Single feature predictions have similar correlation strength with truth as the correlation strength between the feature and mass accretion history.  Larger feature subsets have more prediction power over a wider range of the mass accretion history; using all listed parameters as features to train \texttt{MultiCAM} can predict mass accretion histories with increasing strength towards later times, particularly between $0.8\gtrsim a\gtrsim 0.9$ that has a strong correlation with truth, $\rho_s\approx0.7$.}\label{fig:dspredmah}
\end{figure*}

The left panel of Fig.~\ref{fig:dspredmah} shows the correlation between each dynamical state parameter and the MAH of our sample.
The right panel depicts the predictive power of different feature subsets when using \texttt{MultiCAM} to predict the mass accretion histories of the sample.
We take the absolute values of the correlations for ease of comparison, where dashed lines correspond to negatively correlated parameters.
The banded regions show the 25-75th percentile bootstrap errors in the measurements, drawing 100 samples with replacement.

Substructure mass fraction measured at $a=1$ ($z=0$) correlates the strongest with MAH, with median correlation $|\rho_{\mathrm{sp}}(f_{m, 200c}, m(a))| \geq 0.6$ for $0.6 \leq a \leq 0.95$ ($0.05 \leq z \leq 0.67$).
It is negatively correlated with the MAH, as the introduction of many substructures indicates a major mass accretion event.  Thus, $m(a)$ is expected to stay smaller through much of this time, with a late time increase corresponding to the recent mass accretion.
The mass fraction of the most massive substructure, $f_{m2, 200c}$ is slightly weaker, with moderate correlation with MAH ($|\rho_s(f_{m2, 200c}, m(a))| \geq 0.5$) over a similar window in time.
This observation can be explained by the fact that $f_{m2, 200c}$ excludes less massive substructures, and hence the impact of minor mergers.
Such a result highlights the importance of minor merger events on MAH.

We find the best correlation with the center of mass offset to be $|\rho_s(\Delta_{200c}, m(a))| \geq 0.6$ at late times of $a \approx0.8$ ($z \approx 0.25$).
Again we expect a negative correlation here as a larger offset suggests the onset of a mass accretion event, and hence a smaller $m(a)$ shortly before fast accretion event leading to a larger offset measurement at $a=1$.
As mass falls into the system, the center of mass of the cluster shifts in the direction of the incoming mass, while the density peak remains near the center of the main progenitor.
The virial ratio has a peak correlation $|\rho_s(\eta_{200c}, m(a))| \geq 0.4$ occurring at $a \approx0.9$.
Note, the magnitude gap parameter shows weak correlations throughout all times with a correlation of $|\rho_s(m_{14,3D}, m(a))| \leq 0.3$ (see Appendix \ref{sec:app:m14}).

This result is expected -- MAH leaves a stronger imprint on $f_{m, 200c}$ than $\Delta_{200c}$ (Fig. \ref{fig:mahdscorr}).
$\Delta_{200c}$ is sensitive to the dynamics of the accreted bodies.
As shown in \cite{wang_merger_2025}, an incoming progenitor typically orbits the main dark matter halo before being captured, thus the center of mass may fluctuate.
They demonstrated that this orbit is shorter for more massive systems (up to $M_{\mathrm{vir}} \le 10^{13} M_\odot$ in their analysis), hence we would expect that the $\Delta_{200c}$ signal diminishes quickly ($<0.5t_d$) in our massive cluster sample.
The progenitor also continually loses mass to the main halo as it orbits, further reducing the impact on the center of mass measured.
\cite{wang_merger_2025} also note that the mass and energy loss is more intense for more massive incoming progenitors.
Therefore the $\Delta_{200c}$ merger signature becomes entangled with the regular dynamics of the cluster members, and is weakened.
Contrast this with $f_{m, 200c}$, which is only sensitive to the monotonic decrease in mass of the incoming progenitor.
As noted in Sec.~\ref{sec:methods:mah}, larger values of $f_{m, 200c}$, $\Delta_{200c}$, suggest recent or ongoing merger activity.
This is equivalent to saying that in the recent-past the normalized peak mass must have been small, before getting a 'jump' from the merger.
A larger $f_{m, 200c}$, $\Delta_{200c}$ corresponding to a smaller $m(a)$ leads to a negative Spearman correlation between these DS parameters and MAH.

For all the parameters we note the following common behavior in their correlations with MAH.
Since the present normalized peak mass $m(a=1) = 1$, all correlations converge to 0 at $a = 1$.
Furthermore, since the imprint of a merger only lasts a few dynamical times ($t_d \approx 1.44\,\mathrm{Gyr}$), the strength of the correlations/predictions is lower for all parameters at scales $a \lesssim 0.5$, as this information cannot propagate all the way to what we would calculate at the present day.
In fact, some studies have suggested that we should see this loss of information for mergers as recent as $a \lesssim 0.7$ \cite{nelson_evolution_2012,chen_imprints_2019, gianfagna_study_2022}.

To assess the predictive power of the dynamical state parameters, we train our \texttt{MultiCAM} model on various dynamical state feature subsets, and use these models to predict $m(a)$ at each scale factor, $a$.
To do this, we split the galaxy cluster sample in a 3:1 ratio of train/test split.
The Spearman correlation coefficient is calculated between the predicted $m(a)$ and the true set $m(a)$.  We repeat this 100 times to find a median correlation between predicted and true values at each scale factor and illustrate the 25-75th percentile in shaded bands.
We report the results in the right panel of Fig.~\ref{fig:dspredmah}.
For direct predictive performance for the formation time $a(m=0.5)$, see Fig.~\ref{fig:formation_time}.

Overall we find that information on substructure mass fraction provides most of the constraining power, and we find marginal gains when including $\eta_{200c}$ and $\Delta_{200c}$.  Further, $m_{14}$ adds very little information, so we omit the results when training only on the dynamical state parameters, as the curve traces the black line corresponding to performance of the model trained on all dynamical state parameters and $m_{14}$.  While there is strong predictive power at $a\approx0.8$, these parameters cannot fully characterize the MAH at any epoch.

Note, \texttt{MultiCAM} trained on any single dynamical state parameter (or $m_{14,3D}$), predicts mass fractions that are correlated with the true mass fraction with similar correlation strength as the Spearman correlation curves in the left panel of Fig.~\ref{fig:dspredmah}.
Using \texttt{MultiCAM} to combine parameters generally leads to tighter constraints on MAH predictions; the model trained on all dynamical state parameters produces the best predictions ($\rho_s(m(a)_{\text{test}}, m(a)_{\text{pred}}) \geq 0.7$) at late times ($a \geq 0.8$).

\subsection{Predictive power of morphological features on mass accretion history}\label{sec:results:smpredmah}

\begin{figure*}[ht]
    \centering
    \includegraphics[width=\textwidth]{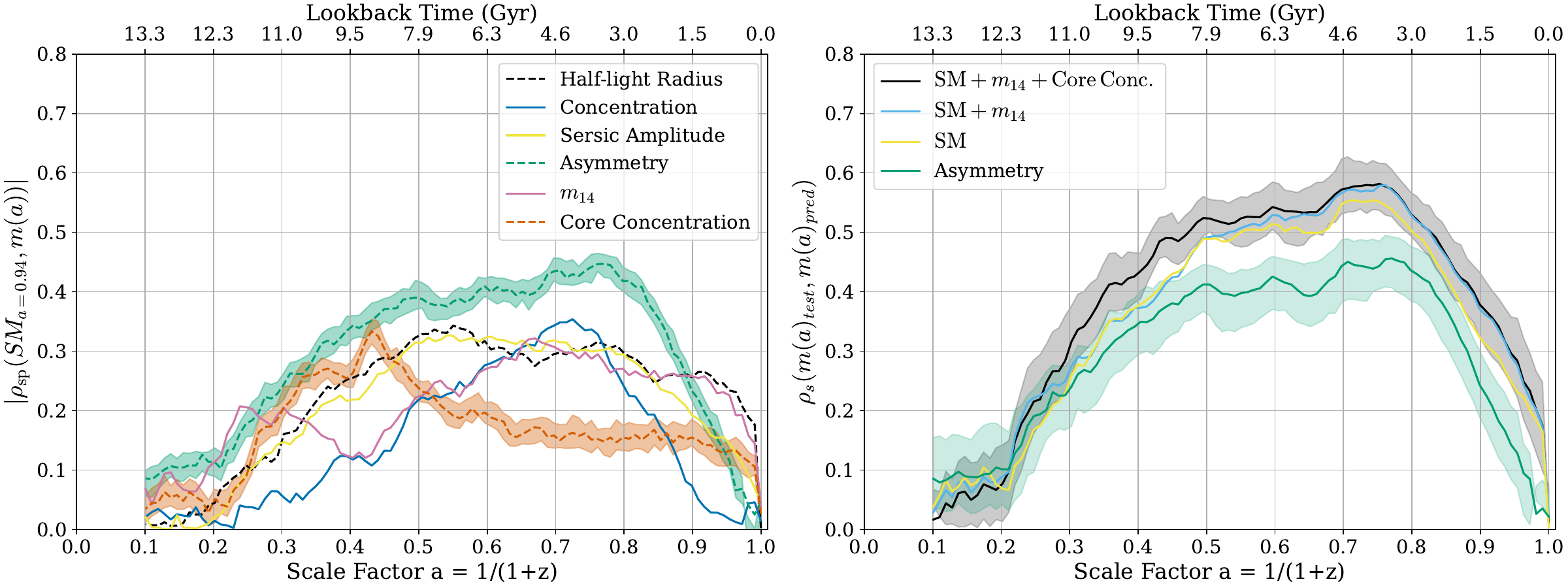}
    \caption{\textit{Left: Correlation between mass accretion history and morphological measurements of the stellar density map}  Solid lines show the Spearman correlation between the mass fraction at a given scale factor, $M(a)/M(a=1) = m(a)$, and various \texttt{statmorph} parameters measured at $a=0.94$, in addition to the magnitude gap $m_{14}$ measured in projection. Shaded regions show example 25th-75th percentile widths of a bootstrap. \textit{Right: Relative prediction power of morphological measurements of the stellar density map with \texttt{MultiCAM}} Solid lines show the Spearman correlation between the the true mass fraction at a given scale factor and mass fraction predicted using different subsets of morphological measurements on the stellar density map as features for \texttt{MultiCAM}. Shaded regions show example 25th-75th percentile widths. Single feature predictions have similar correlation strength with truth as the correlation strength between the feature and mass accretion history.  Larger feature subsets have more prediction power over a wider range of the mass accretion history; using all listed parameters as features to train \texttt{MultiCAM} can predict mass accretion histories between $0.4\gtrsim a\gtrsim 0.85$ that have the strongest correlation with truth, $\rho_s\approx0.5$.}\label{fig:statmorphpredmah}
\end{figure*}

In Fig~\ref{fig:statmorphpredmah} we perform a similar exercise using morphological parameter measurements on projected stellar density maps.
Overall, morphological features of projected stellar density maps show weaker correlations with MAH than dynamical state parameters across all epochs.  We briefly tested the correlation strength of morphological parameter measurements made within different inner-outer radial ranges of the stellar density maps (shown in Fig.~\ref{fig:radconc} and Fig.~\ref{fig:radasym}) and chose several morphological measurements made in core-excised regions within $50~\mathrm{kpc}<r<1~\mathrm{Mpc}$, and a core concentration measured within $30$~kpc.

Asymmetry has a moderate correlation ($\rho_{\text{sp}}(A, m(a)) \gtrapprox 0.4$) found during $0.6 \leq a \leq 0.8$.
Concentration (core-excised) also has a moderate correlation ($\rho_{\text{sp}}(C, m(a)) \sim 0.4$) at late times ($a \approx 0.75$).
Finally, core concentration shows a moderate correlation $\rho_{\mathrm{sp}}(C_{r < 30 \mathrm{kpc}}, m(a)) \approx 0.3$ at early times ($a \approx 0.45$). Note, none of these have correlations as strong as $f_m$, $\delta$, nor $f_{m2}$ at their peak correlation times.  However, these these trends do compare with those of $\eta$ and $m_{14,3D}$ discussed in Sec.~\ref{sec:results:dspredmah}.

We split the galaxy cluster sample into a 3:1 train/test split.
Each cluster has morphological parameter measurements along 3 orthogonal projections, so each cluster appears 3 times in either the training set or the test set (but not both).
We repeat the train/test split 100 times to measure a median and 25-75th percentile Spearman correlation metric of model performance, illustrated with the shaded regions in the right panel of Fig.~\ref{fig:statmorphpredmah}.
Training our MultiCAM model with all the morphologies gives us the strongest predictive power ($\rho_{\text{sp}}(m(a)_{\text{test}}, m(a)_{\text{pred}}) \geq 0.5$) found during $0.5 \leq a \leq 0.8$, reaching a peak $\rho_{\text{sp}}$ slightly above 0.7.
We further find that $m_{14}$ provides complementary late time predictive power ($a \geq 0.5$) when used with our strongest morphological features, $C_{<30~\mathrm{kpc}}$ and $A_{50~\mathrm{kpc},1~\mathrm{Mpc}}$ (blue and yellow comparison in Fig.~\ref{fig:statmorphpredmah}.
These results emphasize the importance of combining multiple morphological parameters to best characterize the MAH.
Any single parameter probe returns moderate-to-weak predictions.

\subsection{Sample Selection by Ranked Mass Accretion History Tracers}\label{sec:results:coeffselectsample}

Several science goals with galaxy cluster samples require either a more or less relaxed subsample selection.
In this section, we assess the potential use of \texttt{MultiCAM} trained coefficients of morphological features in sample selection.
We also compare to raw Spearman correlation coefficients of these features with the MAH.
As a benchmark, we do a similar exercise with dynamical state parameters.

\begin{figure*}[ht]
    \centering
    \includegraphics[width=\textwidth]{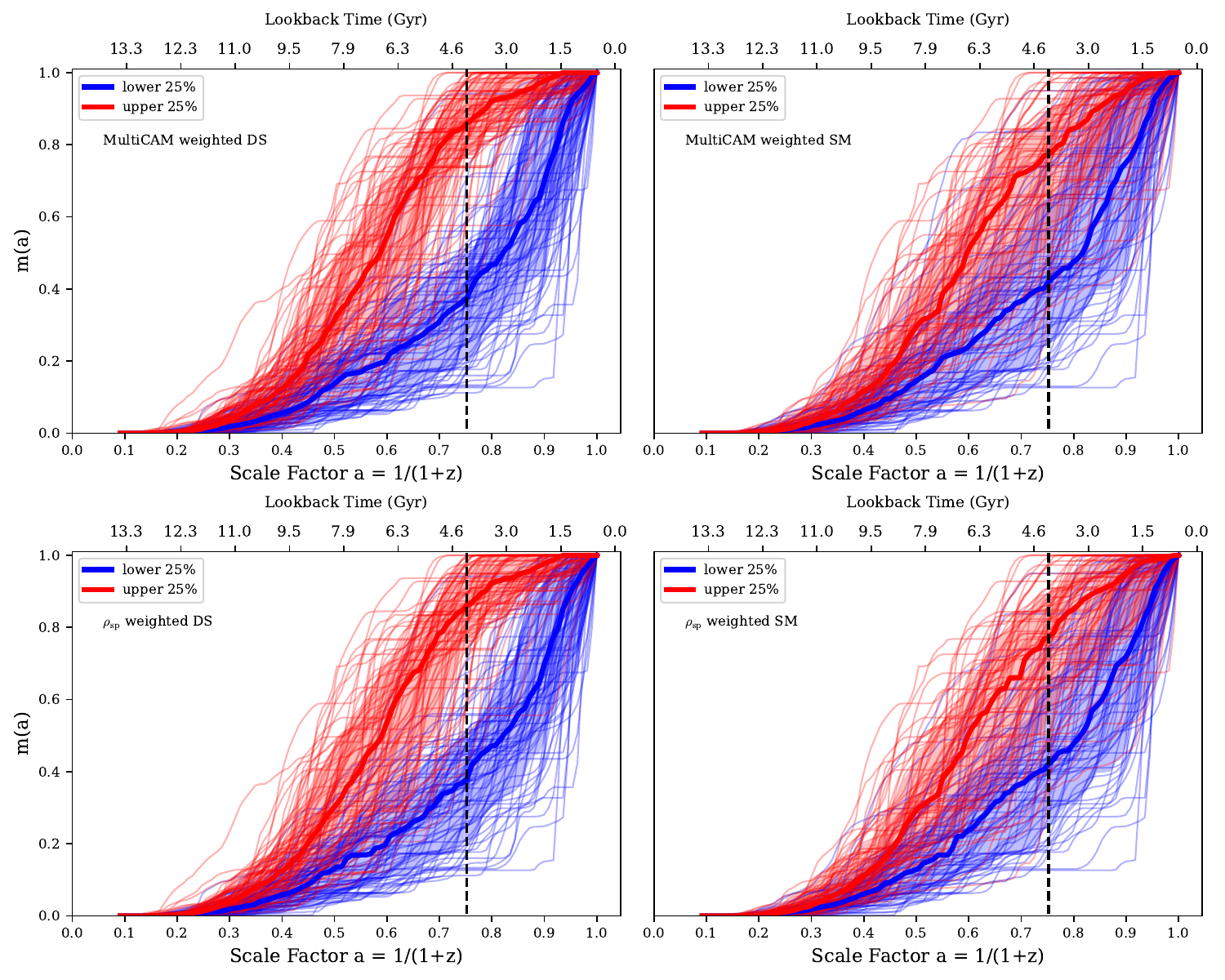}
    \caption{\textit{Selection by ranked mass accretion history indicators.}
     Top row: Ranking using trained MultiCAM coefficients on DS parameters (left) and SM measurements (right).  Bottom row: Ranking using relative Spearman coefficients between mass accretion histories and DS parameters (left) and SM measurements (right). Vertical black dashed line indicates coefficients chosen at $a \approx0.75$ in each panel.
    Highlighted are the median of the lower quartile (blue) and median of the upper quartile (red).
    Contours are the the 16th-84th percentiles of the extreme groups.
    We emphasize the value of this procedure being that this is valid even for imperfect $m(a)$ predictions as rank-ordering is preserved.
    }
    \label{fig:mah_split}
\end{figure*}

We rank order our galaxy cluster sample by their predicted mass fraction at a given scale based on their present-day morphologies and dynamical state parameters in Fig.~\ref{fig:mah_split}).
The results of sections \ref{sec:results:dspredmah} and \ref{sec:results:smpredmah} return MultiCAM's linear regression coefficients and Spearman correlation coefficients for each feature at all times.
We selected the coefficients at a specific epoch ($a=0.752$ / $z=0.33$) as weights to assign scores to the clusters using their present-day features.
The coefficients are chosen at $a=0.752$ as this is an epoch where most parameters show moderate-to-strong correlations with MAH based off of Fig. \ref{fig:dspredmah} and Fig. \ref{fig:statmorphpredmah}.
This weighted sum effectively returns $m(a=0.752)$ for each cluster.
The weights were averaged over 100 Monte Carlo cross-validations for the MultiCAM weights, and averaged over 100 bootstraps for the Spearman coefficients.
We then split the MAH of our sample into upper and lower quartiles.
We quantified the strength of the split by taking the difference in $m(a)$ between the medians of the upper and lower quartiles at $a=0.752$ ($z=0.33$).
The larger the gap, the better the model is at distinguishing between early-forming and late-forming clusters; and so the model is better at characterizing our sample's dynamical history.

When ranking clusters using their MultiCAM coefficients we found that MAH has the clearest gap, therefore enabling an efficient split.
Quantitatively, this split is equal to 0.484 and 0.353 for MultiCAM models trained on dynamical state parameters and morphologies respectively.
When using the spearman correlation coefficients we find comparable results; $\text{split} = 0.480$ and $0.353$ for dynamical state and morphologies respectively.
These results further support our previous findings that the present day morphologies and dynamical states carry imprints of the mass accretion histories of galaxy clusters, and that we can use these parameters to infer the growth of these systems.

\section{Discussion and Conclusions}\label{sec:conlusion}

MAH at any given epoch is not an isolated event: for a limited time it leaves its imprint on the future evolution of the cluster.
At any given time the evolution of the cluster's mass shows tight correlations with the mass accretion in the recent-past, and near-future.
In Fig.~\ref{fig:mahdscorr} we visualize the lifetime of the dynamical imprints for our cluster sample, where we find similar results to \cite{wang_concentrations_2020} that these imprints are detectable over several dynamical times.
However, we take this a step further.
We demonstrate that the time-lag to detecting the imprints of MAH is not constant across all times, meaning the choice of dynamical state parameter will not be universally suitable for clusters detected at different redshifts.
Consider $f_m(a_i = 1)$.
There is a lag of $a\sim0.05$ to where the imprint of $m(a_j)$ is most detectable from $0.65 \lesssim a_j \lesssim 0.95$.
However, for $f(a_i \sim 0.5)$ the lag is even further back in time ($a \sim 0.15$), and is much briefer for a peak signal strength lasting from $0.3 \lesssim a_i \lesssim 0.35$.
This result has important implications for the dynamical state metrics defined in this study.
These metrics are defined using $z=0$ cluster samples, but we have demonstrated that they are less suited to clusters detected at higher redshifts (smaller scale factor).
As more sensitive surveys with more sophisticated cluster detection algorithms increase the sizes of cluster and proto-cluster samples at higher redshifts, this effect must be seriously accounted for when using any observational proxies for dynamical state to constrain the MAH.

In each panel of Fig.~\ref{fig:mahdscorr} there is a blue arc below the diagonal line, indicating a time lag with a systematic anti-correlation between DS parameter measurements occurring some time after the mass measurements (i.e. ${\rho_s}\lesssim-0.2$ where $a_i>a_j$).
The anti-correlation is consistent with dynamically disturbed systems experiencing fast accretion events.
We note that it is unclear if this is a physical feature, artifacts from the simulation, or some 'quirk' associated with the DS parameters used in this study.
Nevertheless, it is interesting that the longest time lag for each feature is around the end of the matter-dominated era of the Universe, $a \approx 0.7$.
This is speculation, more work must be done to explain this.

We demonstrated that we can predict both normalized peak mass, $m(a)$ (see Fig.~\ref{fig:dspredmah}), as well as the time when a cluster first reaches a mass fraction, $a(m)$ (see Fig.~\ref{fig:ds_am_preds}).
These predictions are optimal when using 3-dimensional information (dynamical state parameters), yet there still exists promising constraining power when using projected information (morphological parameters).
We expect that constructing this projected information with multi-wavelength measurements will strengthen our results, which we leave to a future analysis.
Furthermore, \texttt{MultiCAM}'s simple assumption of rank ordering precludes the analysis from much of the problems associated with simulation-driven biases and model misspecification in observations.
Simulations rely on subgrid physics, and the observables extracted here do not necessarily agree with real observations.
The rank-ordering procedure thus helps bridge the gap from simulations to observations.
In other words, the exact values of asymmetry, $m_{14}$ etc. are not necessary -- the rank ordering is sufficient to constrain either the mass fraction at a given time, $m(a)$, or the time when the clusters achieve some fraction of the present-day mass, $a(m)$.
This procedure is not infallible, and could still suffer from challenges from data quality in observations.
Specifically, masking contaminants from non-cluster-member galaxies using photometric/spectroscopic redshift information will be important for extracting robust morphological measurements.
Such contaminants would unpredictably affect the true ranks of the parameter distributions.

Future prospects of the work will be to apply this analysis to observed multi-wavelength datasets to compute the rank ordering of their MAH.
The best proxies for stellar mass are red optical imaging, such as from LSST \citep{ivezic_lsst_2019} or DES \citep{abbott_dark_2016} $i$ and $z$ bands, or near-IR imaging, such as from Euclid \citep{scaramella_euclid_2022}.
Such datasets will be a directly translatable application from the present analysis on stellar density, but with new challenges especially with subtracting background structures before marking our morphological measurements.
Sufficiently deep imaging will be needed to get low surface brightness features in particular, which these surveys may provide.

We will explore how these constraints on MAH affect sample selection at different redshifts in cluster abundance measurements for the most/least formed clusters, allowing improved calibrations of observable-mass scaling relations, and improved constraints on cosmological parameters.
Similar to \cite{haggar_constraining_2024, mpetha_infall_2024} that utilize measurements of the splashback radius, we will explore how these constraints on MAH can facilitate complementary cosmological probes for breaking the $S_8$ degeneracy via cluster formation times.

In this work we quantify the connection between morphological features in stellar density maps and the MAH of a simulated galaxy cluster population, and use the galaxy-halo connection model, \texttt{MultiCAM}, to assess the prediction power of such features on the MAH.
We compare this connection with that from simulation-based dark matter halo features traditionally used to identify relaxed halo subsamples.
We summarize our findings as the following:

\begin{itemize}
\item We measure individual correlations between the MAH of our galaxy cluster sample and individual dynamical state parameters and morphological parameter measurements on projected stellar density maps of the sample.
\item We found that the strongest correlation between the dynamical state parameters and MAH is with the substructure mass fraction ($\rho_{\mathrm{sp}}\approx 0.6$) during $0.5\lesssim a \lesssim 0.9$ ($1 \gtrsim z \gtrsim 0.11$).
The strongest correlation between the morphological features and MAH is with the asymmetry measurement, yielding a moderate correlation ($\rho_{\mathrm{sp}} \approx 0.4$) during later times $0.4\lesssim a \lesssim 0.8$ ($1.5 \gtrsim z \gtrsim 0.25$).
See left panels of Fig.~\ref{fig:dspredmah} and Fig.~\ref{fig:statmorphpredmah}.
\item Our \texttt{MultiCAM} models predict both the normalized peak mass at a given epoch, $m(a)$, and the time when a cluster first reaches a normalized mass fraction, $a(m)$.
\item \texttt{MultiCAM} predictions of the MAH from individual features have correlations with truth that are equivalent to the correlation between that feature and the MAH.
Predictions improve for \texttt{MultiCAM} models trained on all features considered in this study; different features contribute relevant early time information (e.g. from central light concentration) or late time information (e.g. from light asymmetry).
See right panels of Fig.~\ref{fig:dspredmah} and Fig.~\ref{fig:statmorphpredmah} for respective $m(a)$ prediction performance.
See Fig.~\ref{fig:ds_am_preds} and Fig.~\ref{fig:sm_am_preds}) for $a(m)$ predictions.
\item \texttt{MultiCAM} models trained using 3-dimensional information (dynamical state parameters) rather than projected information have better predictive power of the MAH.
Peak correlations between predicted MAH from dynamical state trained models and true MAH reaches $\rho_{\mathrm{sp}}\approx 0.75$ around $a\approx0.85$ ($z\approx0.18$).
On the other hand, peak correlations between MAH predictions from models trained on morphologies and true mass accretion histories reach $\rho_{\mathrm{sp}}\approx 0.55$ around $0.5\lesssim a\lesssim 0.8$ ($1 \gtrsim z \gtrsim 0.25$).
\item We illustrate an example case of sample selection with \texttt{MultiCAM} trained weights, where we rank order the most formed clusters by $a=0.75$ ($z=0.33$).
We compare this to a rank ordering based on measured Spearman correlation coefficients and show improved separation for the \texttt{MultiCAM} trained weights in Fig.~\ref{fig:mah_split})
\end{itemize}

These results show how recent mass accretion leaves its imprints on both the present-day theoretical dynamical state parameters and morphology of the projected stellar density of a galaxy cluster.
We highlight that \texttt{MultiCAM} is a linear model based on a simple assumption of rank ordering.
\texttt{MultiCAM} models trained on a simulation can be applied to an observed sample with a similar initial selection as the simulation with the assumption that the features hold a similar rank ordering.
Thus, even if the shape of the distribution of features differ due to modeling choices, specifically for baryonic feedback, it would be reasonable to apply the model with the assumption that basic rank ordering of features is preserved.

\bibliography{cluster_morphs}
\bibliographystyle{aasjournal}

\begin{appendix}

\section{Appendix}\label{sec:appendix}

\subsection{Regional morphological measurements trace different parts of mass accretion history}\label{sec:app:corevsannuluscorrelation}

\begin{figure*}
    \includegraphics[width=0.9\textwidth]{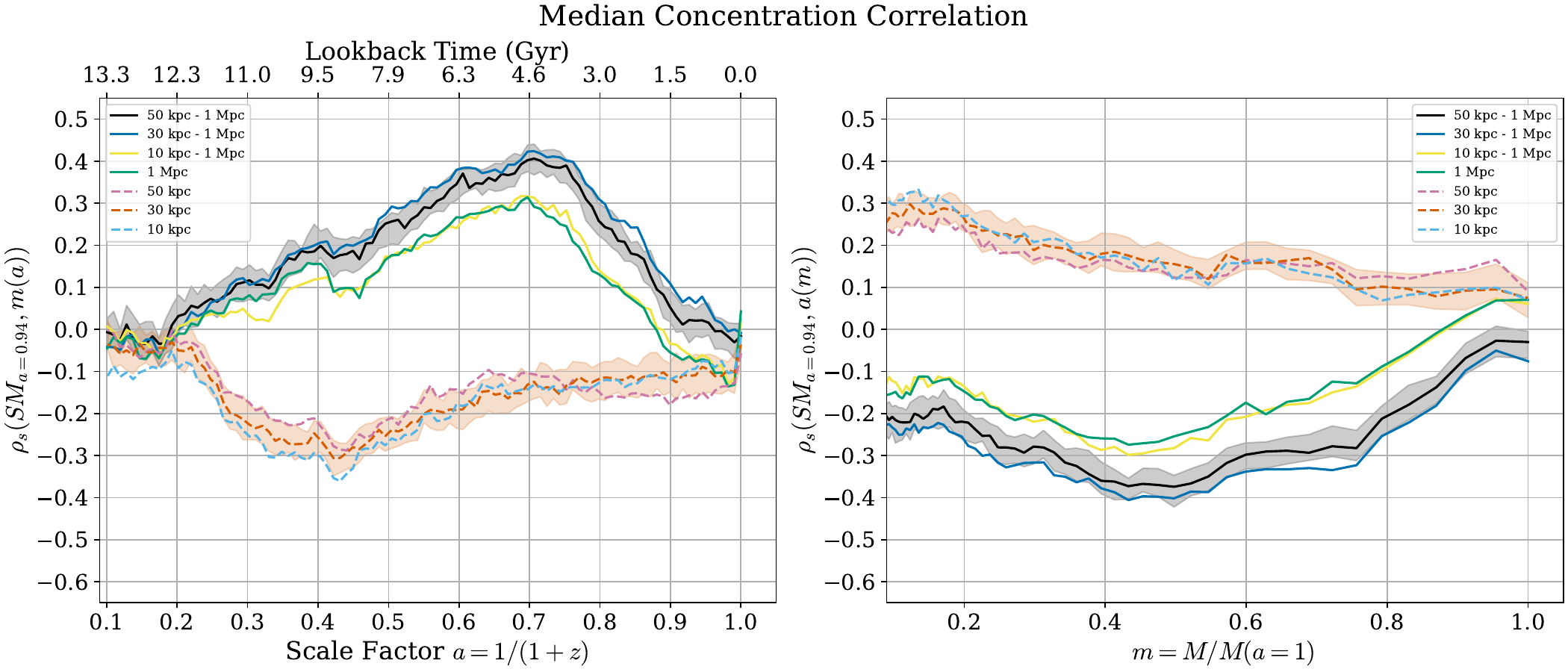}
    \caption{\textit{Correlation between concentration measured in different regions and mass accretion history:}  Dashed lines correspond to correlations between mass accretion history and concentration measured from the cluster core in different apertures. Solid lines correspond to the correlation with core-excised light concentration, measured from some annulus out to 1~Mpc.
    \textit{Left}: MAH parametrized as m(a), the normalized peak mass.
    \textit{Right}: MAH parametrized as a(m), the scale factor when a normalized peak mass is first achieved.
    Colored bands indicate the 25-75th percentiles from jackknife resampling for the $50~\mathrm{kpc} \leq r \leq 1~\mathrm{Mpc}$ region, and the $r \leq 30~\mathrm{kpc}$ region.
    Concentration measured within the core or BCG area better correlates with the mass accreted at early times ($a \leq 0.5$), or when the cluster was less than 20\% of its present-day peak mass ($M/M(z=0) \leq 0.2$).
    The core-excised concentration has the strongest correlation with late time mass fraction ($a \geq 0.6$), or with when the cluster accreted 40-50\% of its present-day peak mass ($0.4\lesssim M/M(a=1) \lesssim0.5$).
    }
    \label{fig:radconc}
\end{figure*}

\begin{figure*}
    \includegraphics[width=0.9\textwidth]{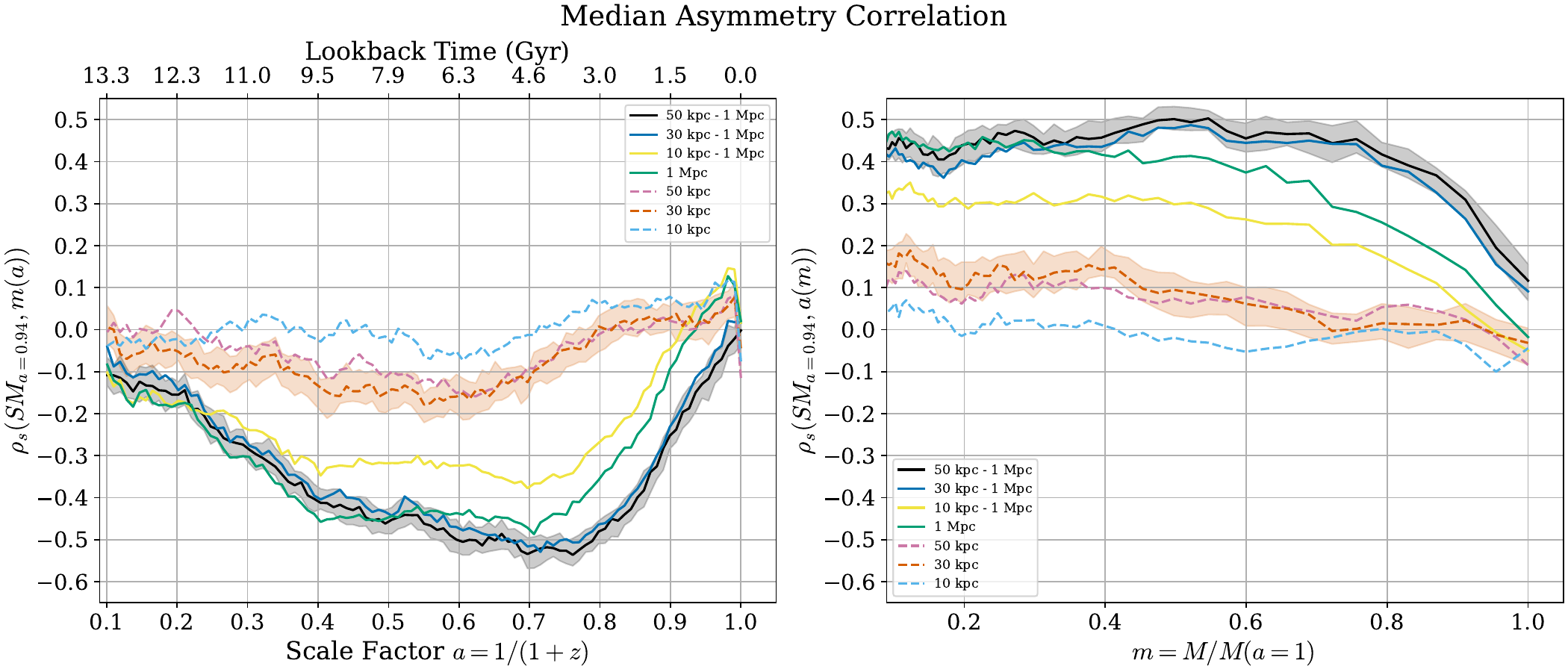}
    \caption{\textit{Correlation between asymmetry in different regions and mass accretion history:} Similar to Fig.~\ref{fig:radconc}, but for the asymmetry parameter.  The core-excised light asymmetry has the strongest correlation with mass fraction at mid-to-late times, $0.5\lesssim a\lesssim0.8$, or with when the cluster accreted 20-80\% of its present-day peak mass.
    }
    \label{fig:radasym}
\end{figure*}

Measurements of the cluster morphology are sensitive to the choice of aperture, and so we explored how our results respond to the aperture used.
We found that a core-excised aperture ($50 \,\mathrm{kpc}\leq r \leq 1 \,\mathrm{Mpc}$) produces the strongest results ($|\rho_{\mathrm{sp}}| \geq 0.4 $) at late times ($a \gtrsim 0.7$) across most of the morphological features, as shown in Fig. \ref{fig:radconc} and Fig. \ref{fig:radasym}.
Furthermore, different parameters respond differently to the choice of the inner radius as can be seen by comparing the strongest correlation curves for the concentration and asymmetry measures in Fig. \ref{fig:radconc} and Fig. \ref{fig:radasym}.
On the other hand, an aperture focused on the cluster core ($r \leq 50\,\mathrm{kpc}$) produces the strongest results ($|\rho_{\mathrm{sp}}| \geq 0.3 $) for the concentration at early times ($a \lesssim 0.5$) where correlations are in general particularly weak.
Alone, this result is not informative, but when combined with the other parameters measured in separate apertures we find consistent improvements in our characterization of the mass accretion history as shown in Fig. \ref{fig:dspredmah} and Fig. \ref{fig:statmorphpredmah}.
This motivates the inclusion of a core concentration, $C_{r < 30 \mathrm{kpc}}$ in the feature vector.
Altogether these tests motivate us to combine different measurements in different regions in the cluster to probe different epochs of the mass accretion history.
Finally on the radius of the aperture, we find that extending the aperture out to $R200c$ needlessly increases computation time while also reducing the strength of the results.
The $1\,\mathrm{Mpc}$ outer radius was found to be an optimal distance for this tradeoff between efficiency and performance.

\subsection{Magnitude Gap}\label{sec:app:m14}

The difference in magnitude between the brightest central galaxy and the fourth brightest cluster member ($m_{14}$) has traditionally been used as a photometric proxy for the hierarchical growth of the BCG, with some evidence suggesting its use as a dynamical state indicator (\cite{golden-marx_hierarchical_2025}).
In this analysis we correlate the magnitude gap with the mass accretion history of our sample of high-mass galaxy clusters.
Previous studies that have found success utilizing $m_{14}$ have been concerned with galaxy groups and low-to-average mass clusters $M \lesssim 10^{14}\,M_{\odot}$ (see \cite{dariush_mass_2010}; \cite{hearin_mind_2013}; \cite{gozaliasl_mining_2014}; \cite{vitorelli_mass_2018}).
At higher masses there is generally a decreased correlation between cluster/halo properties and $m_{14}$ (\cite{farahi_aging_2020}).
It is expected that this also applies to correlations between $m_{14}$ and mass accretion history for high mass systems, such as used in this study ($M_{200c} > 6.42 \times 10^{14} h^{-1}\text{M}_{\odot}$) for the following reason:
More massive systems take longer to form than less massive systems as a consequence of  hierarchical structure formation.
Hierarchical formation preferentially grows the BCG ($m_1$) as dynamical friction sends accreted structures to the halo center.
Satellite members on the other hand mostly maintain their sizes, or are cannibalized by the BCG and so $m_4$ is expected to decrease.
Altogether, this suggests that earlier forming (less massive, more relaxed at $z=0$) systems exhibit larger $m_{14}$ and so we expect a stronger correlation with the normalized peak-mass $m(a)$ in late times.

\begin{figure*}
    \centering\includegraphics[width=\textwidth]{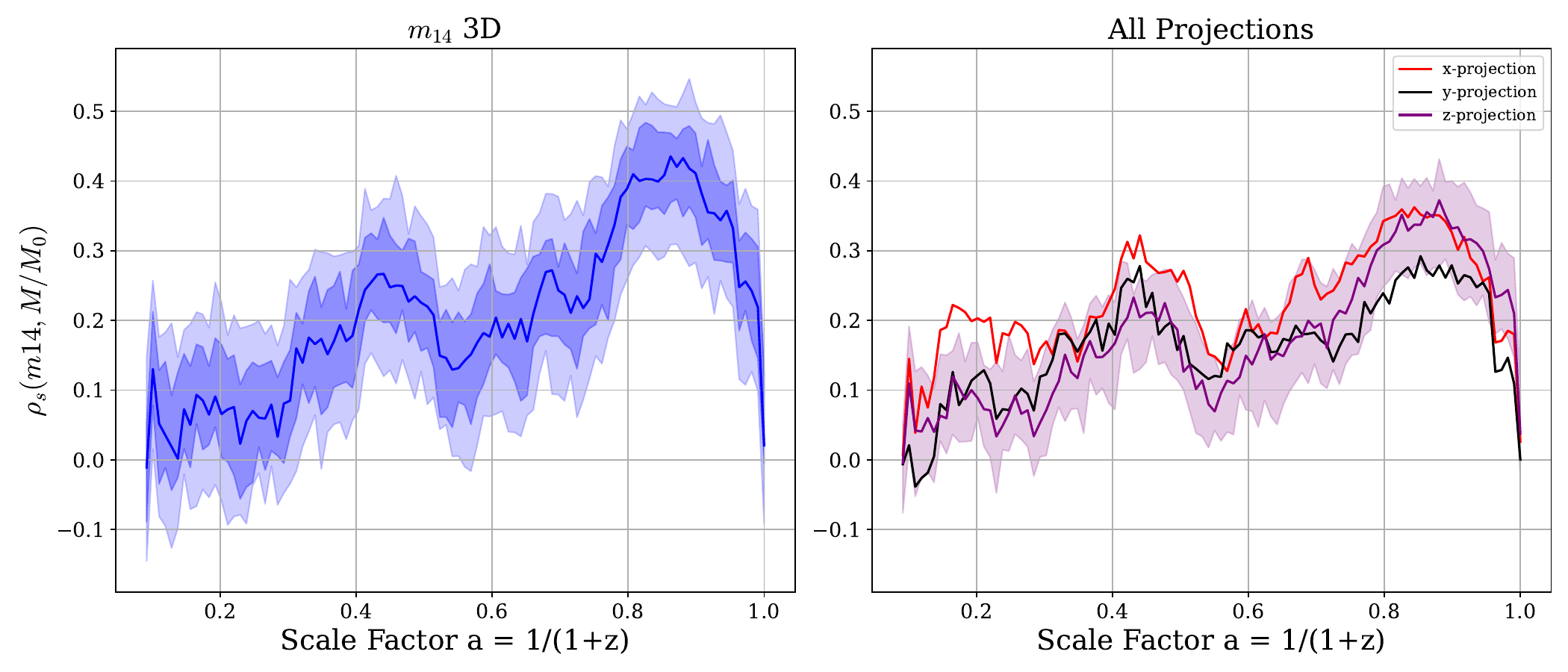}
    \caption{\textit{Correlation between magnitude gap, $m_{14}$, and mass accretion histories:}  We subselect 147 clusters with the largest $m_{14,3D}$ (earliest forming objects), similar to the selection from \cite{casas_optical_2024} to illustrate the stronger dependence. This is consistent with findings in \cite{golden-marx_hierarchical_2025} that clusters with larger $M_{14}$ form earlier, and are more relaxed at $z=1$.
    Colored bands indicate the 25-75th from jackknife resampling.  Left panel shows the correlation for $M_{14}$ measured from 3D simulation data.  Right panel shows the correlation for $M_{14}$ measured from projected stellar maps.  Peak correlation occurs between $M_{14}$ and mass accreted at $a\approx0.85$.}
    \label{fig:m14_mah_corr}
\end{figure*}

To test this we first subselect systems with $m_{14} > 1.25$ and correlate $m_{14}$ with $m(a)$.
As shown in Fig. \ref{fig:m14_mah_corr} we find improved median correlative strength $\rho_{\mathrm{sp}} \geq 0.4$ at late times, $a > 0.8$, as compared to using the full sample (see Fig. \ref{fig:statmorphpredmah}).
Furthermore, the signal is comparable to our best morphological measurement, the asymmetry, when we apply this cut.
To make our results more comparable to previous analyses (although we are still in the high-mass regime), we also selected the bottom half of our clusters in mass to test how $m_{14}$ responds.
With this mass-cut we find similar results ($\rho_{\mathrm{sp}} \geq 0.4$) to applying the $m_{14}$ cut, albeit at later times of $a \sim 0.9$. We omit these curves as they show the same shape for the majority of the accretion history, just peaking later.

These results support the theory expectations for the magnitude gap in high mass systems.
They also explain why the magnitude gap does not perform as well in predicting mass accretion histories and formation times in our cluster sample as compared to other studies.

\begin{figure*}[ht]
    \centering
    \includegraphics[width=\textwidth]{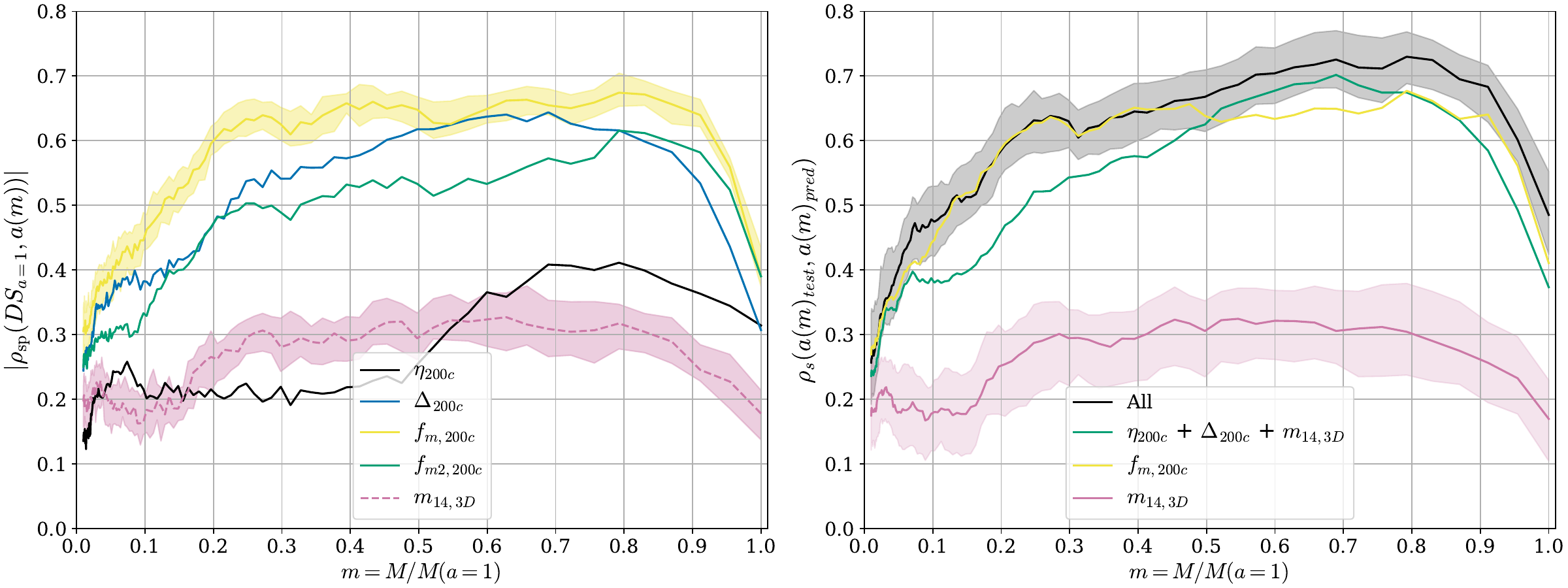}
    \caption{\textit{Left: Correlation between mass accretion history and traditional dynamical state parameters}. Solid lines show the Spearman correlation between $a(m)$ and various dynamical state parameters from the dark matter particles measured at $a=1$, in addition to the magnitude gap $m_{14}$ measured in 3-D. Dashed lines show parameters that negatively correlate with $a(m)$. Shaded regions show example 25th-75th percentile widths of a bootstrap.  \textit{Right: Relative prediction power of traditional dynamical state parameters with \texttt{MultiCAM}}. Solid lines show the Spearman correlation between the the true and predicted $a(m)$ using different subsets of dynamical state parameter measurements as features for \texttt{MultiCAM}. Shaded regions show example 25th-75th percentile widths.}\label{fig:ds_am_preds}
\end{figure*}

\begin{figure*}[ht]
    \centering
    \includegraphics[width=\textwidth]{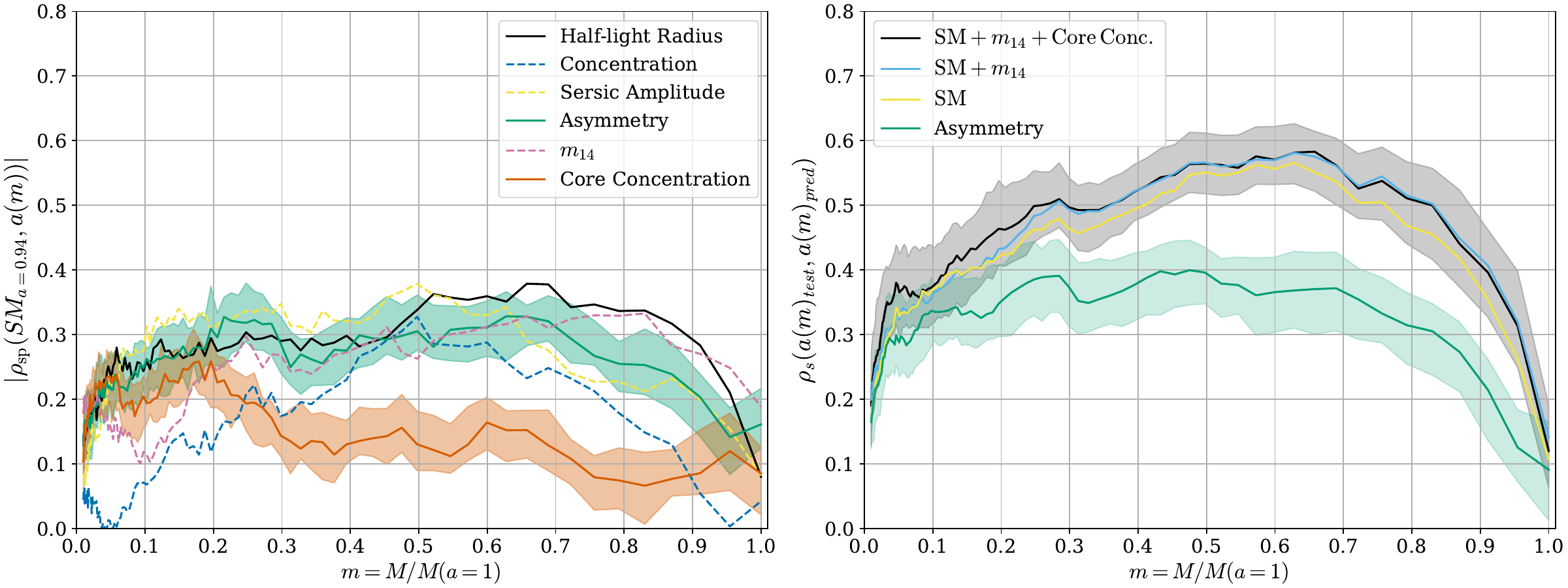}
    \caption{Same plot as Fig.~\ref{fig:ds_am_preds}, but with the morphological measurements in projection, measured at $a=0.94$.}\label{fig:sm_am_preds}
\end{figure*}

\subsection{Predicting More Proxies for Mass Accretion History}\label{sec:app:mah}

The normalized peak mass, $m(a) = M_{\mathrm{peak}}(a) / M_{\mathrm{peak}}(a=1)$ is just one of a number of proxies for the mass accretion history.
We can reparameterize MAH to be $a(m)$, the first time that a cluster reaches a fraction of its present-day mass.
To do this, we take advantage of the monotonicity of $m(a)$ (see Sec. \ref{sec:methods:mah} for details) and simply invert the function $m(a)^{-1} = a(m)$.

We present the results connecting $a(m)$ to the dynamical state in the left panel of Fig. \ref{fig:ds_am_preds}.
We find the strongest correlation in the substructure mass fraction $\rho_{\mathrm{sp}} \geq 0.6$ over a broad range of peak masses ($0.2 \leq m \leq 0.95$).
There are similarly broad, but weaker, correlations in the other parameters.
The reason for this can be seen in the peak correlations in the left panel of \ref{fig:dspredmah}.
The correlation in $f_m$ approaches its peak around $a \sim 0.5$, and if we look at the distribution of $m(a=0.5)$ in Fig. \ref{fig:mah_split} we see a broad range of normalized masses in our sample.
Conversely, at $a \leq 0.2$ and $a \geq 0.95$ the $m(a)$ distribution is much narrower and so there are a wide range of $f_m$ assigned to a small range in $m(a)$, hence the correlations over these time frames in Fig. \ref{fig:ds_am_preds} are weak.

In the right panel of fig \ref{fig:ds_am_preds} we show the correlations between MultiCAM predicted and the true $a(m)$.
We find the strongest correlation ($\rho_{\mathrm{sp}} \geq 0.7$) for our MultiCAM model trained on the full dynamical state feature vector for $m \geq 0.6$.
The result is comparable to the peak correlation when using $m(a)$ in Fig. \ref{fig:dspredmah}.

We repeat the above procedure using the \texttt{statmorph} morphologies as shown in Fig. \ref{fig:sm_am_preds}.
The same caveats apply to how we interpret the breadth of the peak correlations.
We emphasize that the strengths of the peak correlations across these two MAH parameterizations are comparable as we are encoding the same information, just in different ways.
This highlights the flexibility in applying these results depending on the context of the analysis.

To further exemplify the flexibility of this analysis we predict a third parametrization of the cluster growth history -- the formation time (see \cite{li_halo_2008} for range of formation time descriptions).
The formation time has been found to be correlated with halo concentration (\cite{navarro_universal_1997}, \cite{wechsler_concentrations_2002}, \cite{bullock_profiles_2001}), as well as the bias in galaxy clustering ( \cite{haggar_constraining_2024, mpetha_infall_2024}, \cite{li_halo_2008}, \cite{gao_age_2005}).
More specifically, it has been found that for cosmologies of varying $[\Omega_M, \sigma_8]$ the average time at which large-scale structure collapses and clusters also varies.
A low $\Omega_M$ means a larger $\Omega_\Lambda$.
Consequently the matter-dominated era of the Universe, the peak of large-scale structure formation, does not last as long -- cluster's will have earlier formation times.
Similarly, a larger amplitude of density fluctuations leads to faster collapse of large-scale structure -- cluster's will form earlier.

\begin{figure*}[tbp]
    \includegraphics[width=0.9\textwidth]{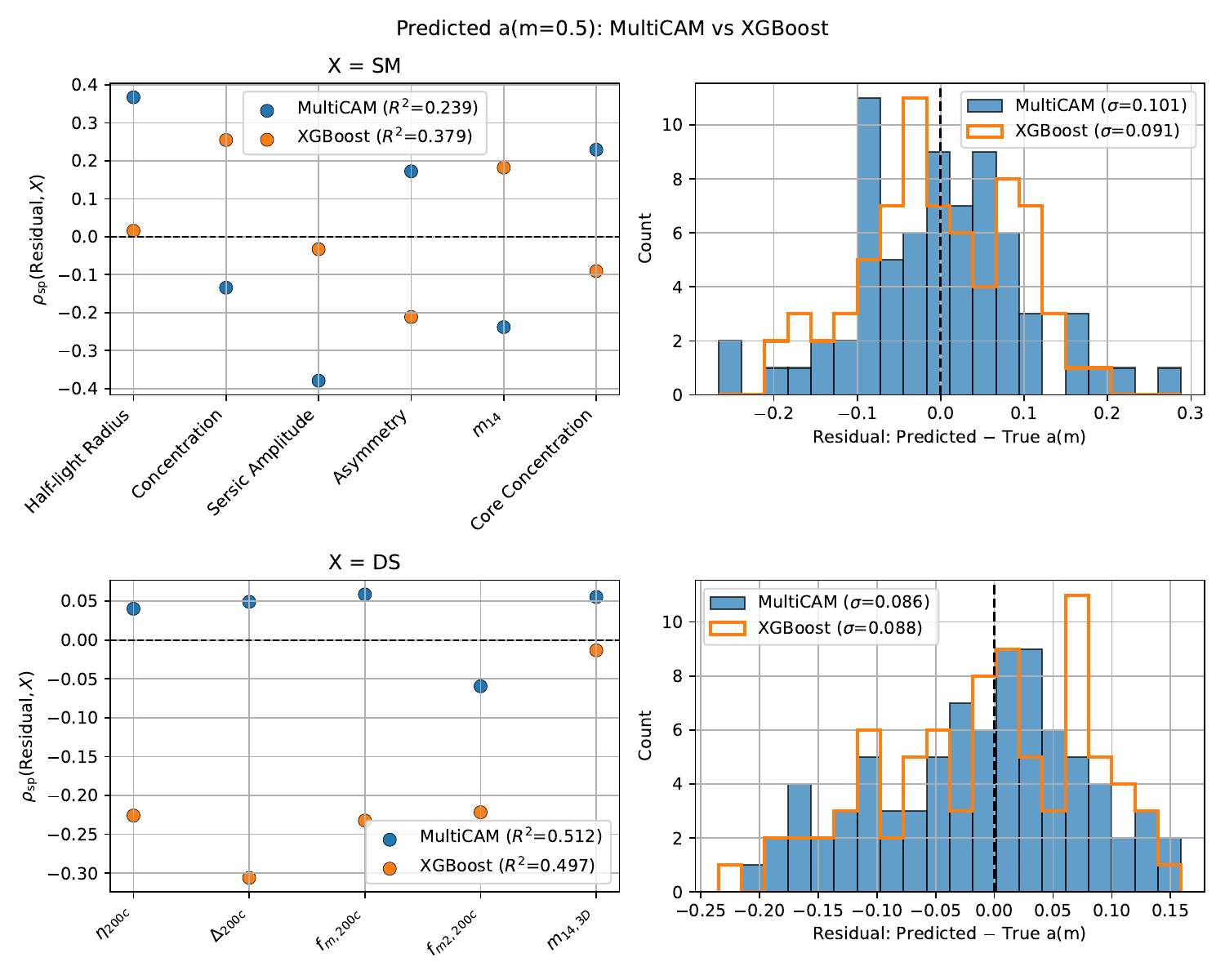}
    \caption{\textit{Predicted cluster formation times.}
    Top: MultiCAM and \texttt{XGBoost} models trained on single-projection \texttt{statmorph} morphologies. Bottom: Models trained on dynamical state parameters.    }
    \label{fig:formation_time}
\end{figure*}

We train our MultiCAM model on morphologies and \cite{neto_statistics_2007} parameters separately, and present predicted formation times in Fig. \ref{fig:formation_time}.
The \texttt{statmorph} model captures 23.9\% of the variation in predicted formation time, and the \cite{neto_statistics_2007} model captures 51.2\%.
We also correlate the residuals with each feature to determine which features the model uses effectively to get predictions.
For the model trained on $DS$ parameters, we find the correlation $|\rho_{\mathrm{sp}}(\mathrm{residuals}, X_i)| \leq 0.05$ which we expect for a model with a good $R^2 = 0.512$.
For the model trained on $SM$ parameters, we find a maximum correlation of $|\rho_{\mathrm{sp}}(\mathrm{residuals}, X_i)| \approx 0.4$.
The latter result is also to be expected for models that cannot capture the variance in the predictions well ($R^2 = 0.239$), but here we show where the model is failing on a feature-by-feature basis.

That so much of the targets' variance is left unaccounted for prompted a brief comparison to a flexible \texttt{XGBoost} \citep{chen_xgboost_2016} regression model.
We tune hyperparameters using the \texttt{GridSearchCV} from the \texttt{SCIKIT-LEARN} \citep{pedregosa_scikit-learn_2018} python package.
Hyperparameters include the learning rate, the number of trees, and the maximum tree depth.
When training on the \cite{neto_statistics_2007} parameters we find a decrease in performance to $\mathrm{R}^2 = 49.7$.
Whereas training on morphologies yields an improvement to $\mathrm{R}^2 = 0.379$.
Similar to the results above, \texttt{XGBoost} models with a higher $R^2$ show worse correlations between residuals and each feature.
When training on $DS$ parameters, it is interesting that although the MultiCAM and \texttt{XGBoost} model have comparable $R^2$ values (0.512 and 0.497 respectively), the MultiCAM model is extracting noticeably more information from the features as-is.
Furthermore, the signs on the correlations are flipped for most features between the MultiCAM and \texttt{XGBoost} model.
For the MultiCAM model, the sign is physically reasonable for each feature (except core concentration): more disturbed systems are harder to predict for, and thus have larger residuals.
These findings highlight the strengths of the MultiCAM model.
Despite its simple rank-ordering principles, it may perform comparably well to complicated non-linear models while maintaining  physically interpretable results.
This will be particularly desirable when translating to observed datasets.

\begin{figure*}[tbp]
    \includegraphics[width=0.9\textwidth]{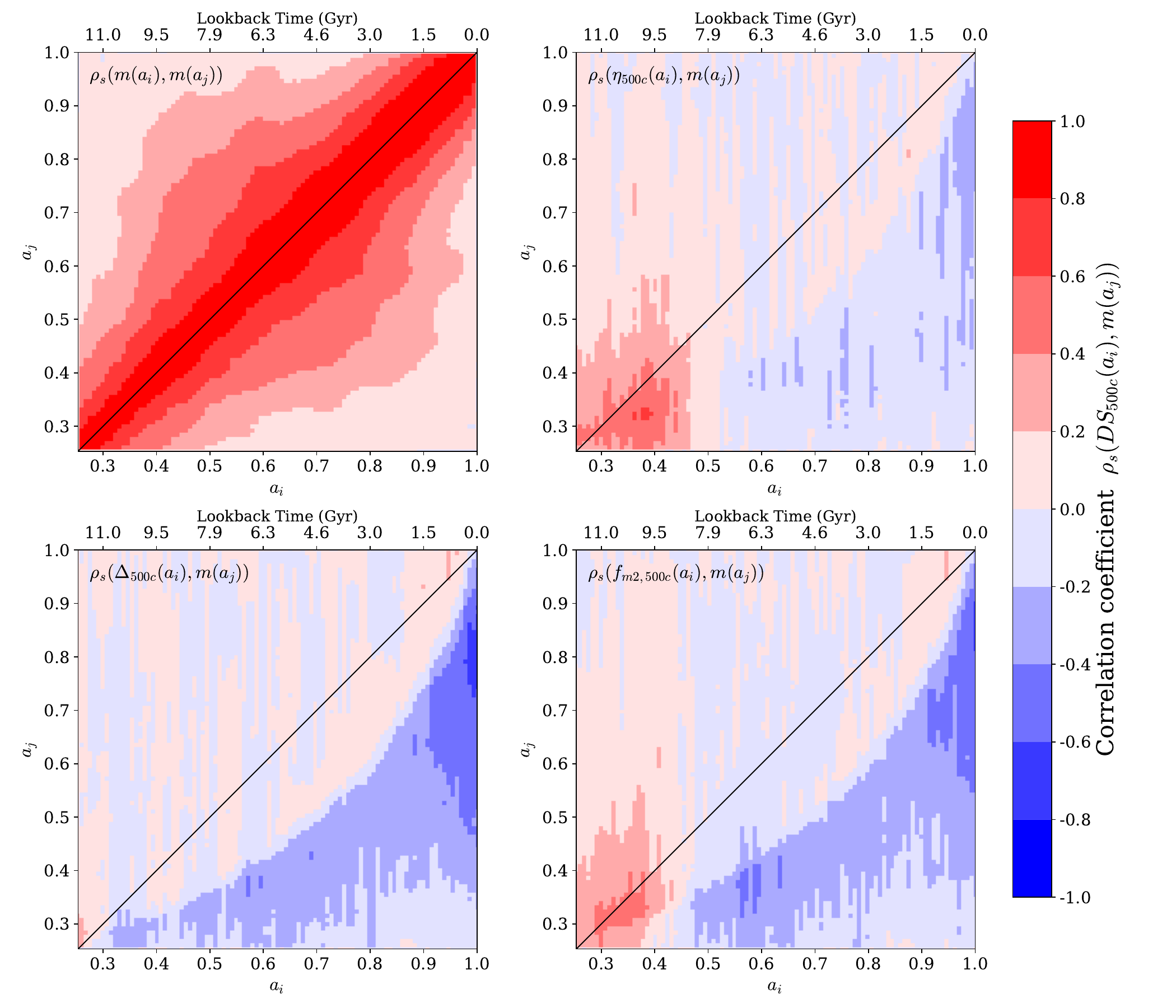}
    \caption{\textit{Correlation between the mass accretion history, $m(a_i)$, and the evolution of traditional dynamical state parameters measured at $R_{500c}$, $DS_{500c}(a_j)$.}
    The colors in each pixel represent the Spearman correlation between $DS_{500c}(a_i)$ and $m(a_j)$, each at times $a_i$ and $a_j$ respectively, with the exception of the top left panel showing the Spearman correlation between mass accreted at different times, $m(a_i)$ and $m(a_j)$.
    The diagonal, over-plotted in black, corresponds to measurements taken at the same time, $a_i = a_j = a$. Traditional dynamical state parameter measurements at $R_{500c}$ retain less information of mass accreted, as compared with measurements at $R_{200c}$ (see Fig. \ref{fig:mahdscorr}).
    In general, the peak signal strength lasts less time when measuring at $R_{500c}$.
}\label{fig:mahdscorr500c}
\end{figure*}

\end{appendix}

\end{document}